\documentclass[%
aip,
amsmath,amssymb,
reprint,%
]{revtex4-1}

\usepackage{graphicx}
\usepackage{dcolumn}
\usepackage{bm}
\usepackage{mathrsfs}

\usepackage[utf8]{inputenc}
\usepackage[T1]{fontenc}
\usepackage{mathptmx}
\usepackage{etoolbox}
\usepackage{lipsum} 
\usepackage{xcolor}
\usepackage{algorithm}
\usepackage{algpseudocode}
\usepackage[most]{tcolorbox}
\usepackage{enumitem}
\tcbuselibrary{listingsutf8}
\usepackage{booktabs} 
\usepackage{bm}
\usepackage{amsmath}
\usepackage{tcolorbox}
\tcbuselibrary{listings, breakable, skins, theorems}

\usepackage{siunitx}  
\makeatletter
\def\@email#1#2{%
	\endgroup
	\patchcmd{\titleblock@produce}
	{\frontmatter@RRAPformat}
	{\frontmatter@RRAPformat{\produce@RRAP{*#1\href{mailto:#2}{#2}}}\frontmatter@RRAPformat}
	{}{}
}%
\makeatother
\begin{document}
	
	\preprint{AIP/123-QED}
	
	\title[\textbf{Phys. Fluids (2025)} | Manuscript under preparation]{Assessment of modern shock capturing schemes for all-speed flows in the OpenFOAM framework}
	\affiliation{ 
		Department of Aerospace Engineering, Indian Institute of Technology Kanpur, 208016, Kanpur, India }

	\author{Anurag Adityanarayan Ray $^{1}$}
        \author{Sreejita Bhaduri $^{1}$}
        \author{Swetarka Das $^{1}$}

	\author{\ Ashoke De $^{1}$}
	\email{ashoke@iitk.ac.in}
	
	\date{\today}
	
	\begin{abstract} 
		
		OpenFOAM is a widely used CFD (Computational Fluid Dynamics) software that relies on the finite volume method formulation for solving a wide range of fluid flow problems. However, the default numerical schemes, especially for shock-capturing methods, e.g., Kurganov, Noelle Petrova (KNP), are lower-order accurate. Several higher-order accurate schemes employing the Flux Vector Splitting Method and approximate Riemann solvers are established in high-fidelity computations. This article demonstrates the implementation of modern Riemann solvers along with AUSM+up (Advection Upstream Splitting Method) and LDFSS (Low Diffusion Flux Splitting Scheme) in the OpenFOAM framework and assesses them by solving various problems of increasing complexity. These numerical experiments show that the default scheme in OpenFOAM is robust, but it’s very diffusive on coarse grids, leading to suppressed flow features, while increasing the grid size generates spurious unphysical oscillations. This solver is stable only under a low Courant number limit, but can sustain these numerical noises at a high Courant number of 0.5. We show that using a simple TVD-based (Total Variation Diminishing) Runge-Kutta time integration method makes the predictor-corrector approach robust and, at the same time, prevents these numerical instabilities. Further, using HLLC (Harten-Lax-van Leer Contact) and its corrected versions [HLLC-LM (Harten Lax van Leer Contact with Low Mach number correction) and HLLCP (Harten Lax van Leer Contact with pressure dissipation)], AUSM+up and LDFSS schemes suppress these spurious oscillations and at the same time, increase the accuracy of the solver, thus enabling superior resolution of shock and contact wave discontinuities even on a coarse grid. We demonstrate that the standard HLLC scheme suffers from grid-aligned shock discontinuities for some of the test cases; however, they are cured when using the corrected versions of this scheme. The AUSM+up shows a slightly elevated amount of numerical dissipation compared to these approximate Riemann solvers, which fail to capture intrinsic instabilities over the contact surfaces, and it underperforms in the deep subsonic flow regime. In contrast, LDFSS shows better characteristics and produces comparable results with HLLC-type schemes; however, the computations are expensive in terms of the number of iterations while considering very low Mach number flows in the incompressible limit. Furthermore, the LDFSS scheme fails for strong shock unsteady propagating shock waves. The key findings in this article will help an OpenFOAM user implement these schemes and make an informed choice regarding selecting an appropriate shock-capturing scheme according to the type of problem being solved.
	\end{abstract}
	
	\maketitle

	\section{\label{Introduction} Introduction and Backgroud}
	High-speed fluid flow problems have been extensively studied for decades through the use of computational and experimental methods in conjunction with analytical theories. The recent advances, primarily in measurement techniques such as PIV (Particle Image Velocimetry), LDV (Laser Doppler Velocimetry), Schlieren, and shadowgraph flow visualisation methods, to name a few, have significantly contributed to the knowledge base of fluid kinematics and its dynamic behaviour that we know to date. Notwithstanding the versatility of these techniques, the experiments typically face difficult obstacles, especially when it comes to the meticulous planning of the experimental setups and, to a certain degree, guaranteeing the perfect, controlled atmosphere in which the tests are anticipated to be carried out. In such controlled settings, these tests are also expected to be very repeatable. It can be extremely time-consuming to ensure it, particularly if the test matrix for the issue being studied is too big. Additionally, the financial outlay for a parametric study of this magnitude can be somewhat high.

    The numerical solutions of the well-defined governing equations of the fluid flow problem have been realized in the modern 21st century, thanks to the significant increase in computational power provided by high-performance computing and GPUs. These solutions have been shown to be successful in reproducing the real fluid flow phenomena, even for large-scale problems. Nowadays, many researchers use this tool/method to help with the tests, circumventing the problems and difficulties indicated above. In order to establish verification and validation of the numerical methods used, it is crucial to compare them with the results of the experimental measurements first. As a result, they can be employed further with confidence and without being overly concerned about the difficulties that come with experimental studies, as previously mentioned.

    The Finite Difference Method (FDM), Finite Volume Method (FVM), and Finite Element Method (FEM)\cite{versteeg2007introduction}
    are just a few of the numerical techniques that have been developed and refined over decades to discretize the governing equations (conversion to algebraic form). Each of them has advantages and disadvantages, but FVM has gained significant traction in the computational world for extracting concepts based on the conservation of quantities principle \cite{versteeg2007introduction, anderson1995computational}. 
    The FVM approach is used to discretize the governing equations in a variety of commercial codes and tools, including Fluent \cite{manual2009ansys},
    Star CCM+, OpenFOAM \cite{medina2015open}, and even in-house codes like PraVaHa and SU2 \cite{economon2016su2}. These codes/software are considered versatile in solving fluid flow problems, covering a spectrum of low-speed incompressible flow to high-speed supersonic and hypersonic flow regimes.

    Some of the above-mentioned commercial software can be somewhat pricey. As an alternative software, OpenFOAM offers a vast collection of massively parallel open-source codes under the GNU Public License. These algorithms, such as $\it{rhoCentralFoam}$ \cite{greenshields2010implementation}, a well-known standard density-based solver for simulating compressible flow, are fairly resilient and versatile, but their lower order of accuracy is limited to $\it{O(2)}$. Furthermore, even with first-order Euler time integration, the usual methods used to resolve strong shock waves result in false oscillations behind them \cite{kumar2021modes, kumar2021role}. On the other hand, the diffusive nature of this solver suppresses the desired flow features that are necessary for high-fidelity numerical simulations, as the authors will show in this study. Numerous studies that have illustrated these limits have focused on this \cite{jiang2022development}.

    In this viewpoint, there is an open scope for implementing different shock-capturing schemes in the OpenFOAM framework that are robust and, simultaneously, higher-order accurate. Therefore, the central theme of this article is to assess the compressible fluid flow problems of varying degrees of complexity and employing various classes of shock-capturing schemes essential for convective flux discretization. By default, two shock-capturing schemes (in OpenFOAM) are widely used, i.e. the central upwind scheme of Kurganov, Noelle, Petrova (KNP)  \cite{kurganov2001semidiscrete}\& central scheme of Kurganov Tadmor (KT) \cite{kurganov2000new}, which belong to the category of central schemes with added dissipation \cite{blazek2015computational}. Besides this, other shock-capturing methods can be bracketed primarily under the Flux Difference Splitting method \& Flux Vector Splitting method \cite{blazek2015computational,toro2013riemann}. Famous Riemann solvers such as HLL \cite{davis1988simplified}, HLLE \cite{einfeldt1988godunov}, HLLC \cite{toro1994restoration}, and Roe \cite{roe1981approximate} fall under the category of the Flux Difference Splitting Method, while Van Leer \cite{van2005flux}, AUSM family schemes \cite{liou1996sequel, liou2006sequel}, \& LDFSS \cite{edwards1997low}  family schemes belong to the Flux Vector Splitting Method. They have excellent dissipation and dispersion characteristics and are usually employed in high-fidelity compressible flow simulations; however, the original forms of these schemes, especially the Riemann solvers, suffer from various numerical anomalies/defects, as identified by Quirk \cite{quirk1997contribution}. Several cures are suggested, e.g. Ref.  \cite{kim2003cures} for the Roe Scheme and Ref.  \cite{xie2019accurate,fleischmann2020shock} corresponding to the HLLC type schemes. They are proven effective in alleviating the defects inherent in the original forms.

    To the best of the author’s knowledge, no studies compare these contemporary schemes under different categories for the numerical dispersion and dissipation characteristics. A part of the novelty of this article is in providing these perspectives by implementing the HLLC, and the corrected versions of HLLC, i.e., HLLCP \cite{xie2019accurate}  and HLLC-LM \cite{fleischmann2020shock}, along with flux vector splitting methods, i.e., LDFSS and AUSM+up, and applying them to the problems of inviscid and viscous nature exposed to extremely low Mach number flow in the incompressible limit to hypersonic range. We also use these comparisons to highlight the defects of OpenFOAM's default solver $(\it{rhoCentralFoam})$, (the only density-based solver that is currently available to address compressible flow problems) when it comes to solving problems on the highly refined grid and/or with a higher CFL number at its stability boundary (CFL = 0.5), and demonstrate that it presents difficulties for the solution's convergence. It is worth noting that the use of the predictor-corrector approach with a non-TVD-based Euler time integration method directly results in numerical instability, as the temperature field update lags due to the sequential solution of the Euler equations. The state variables should ideally be updated simultaneously. However, in the \textit{rhoCentralFoam} solver, the temperature is calculated from the total internal energy after updating the velocity field at the current time step, resulting in the incorrect evaluation of the numerical speed of sound and the corresponding interface Mach number on which the convective flux discretization relies. As a result, the error accumulated with this lag increases with each iteration, and it is much higher for the fine grid and/or high CFL numbers where the employed scheme's (e.g. convective flux discretization scheme) inherent artificial viscosity is unable to suppress it. 

Therefore, this article's novelty is also partly associated with implementing the TVD-based Runge-Kutta integration scheme and improving the accuracy and robustness of the default predictor-corrector approach that relies on first-order Euler integration. This superior algorithm boosts the accuracy while simultaneously maintaining the robustness of the algorithm, even when using a CFL number of 0.5 on a dense grid, irrespective of the implemented schemes used. In a sense, this paper offers users interested in high-fidelity robust simulations in the OpenFOAM framework a comprehensive guide that aids in their decision-making about shock-capturing strategies. 

The remainder of the paper is structured as follows following this succinct introduction: Section \ref{Section2} provides the necessary formulation of the Navier-Stokes equation including the detailed aspects of the aforementioned schemes; Section \ref{Section3} provides results and comparative analysis between different flux schemes along with some of the standard solver's solution in the OpenFOAM framework by solving problems of one, two and three-dimensional nature across the extreme ends of Mach number (incompressibe to hypersonic regime of flow) in the Euler framework where we do not consider the effects of physical viscosity; Section \ref{Section4} extends this investigation by imposing the effects of physical viscosity by solving the Navier-Stokes equation for the two dimensional cases. At last, Section \ref{Section5} summarises the key findings of this numerical investigation and provides concrete conclusions.

\section{Numerical methodology \label{Section2}} 
This section of the article elaborately discusses the various contemporary numerical schemes for the convective flux calculation. The intention of the authors in this section is to comprehensively provide all the necessary formulations for the different schemes that would help the reader to implement them accordingly to their use. We try to refrain from changing the notations as far as possible, as described in the respective work, for the ease of understanding.

\subsection{Governing equations}

The Navier-Stokes equations are solved in the OpenFOAM framework, representing mass, momentum and energy conservation equations, and they are listed below in dimensional form.
	
	\begin{eqnarray}
		\frac{\partial \rho}{\partial t}+\frac{\partial (\rho u_j)}{\partial x_j} = 0\;,
		\label{E1}
		\\
		\frac{\partial (\rho u_i)}{\partial t}+\frac{\partial (\rho u_i u_j)}{\partial x_j} = -\frac{\partial p}{\partial x_i} + \frac{\partial \tau_{ij}}{\partial x_j}\;,
		\label{E2}
		\\
		\frac{\partial (\rho E)}{\partial t}+\frac{\partial (\rho E u_j)}{\partial x_j}+\frac{\partial p u_j}{\partial x_j} = \frac{\partial \tau_{ij} u_i}{\partial x_j}-\frac{\partial q_j}{\partial x_j}
		\label{E3}.
	\end{eqnarray}
	
Here, $\rho, u_{i}, u_{j}, p, \tau_{ij},$ and $E$ represent the various flow properties corresponding to density, velocity components, pressure, viscous shear stresses, and total energy. $E$ is the total energy of the flow comprising of internal energy ($e$) and the kinetic energy, i.e., $E=e+0.5u_j^2$. Further, the viscous shear stress $\tau_{ij}$ and heat transfer rate $q_j$ can be expressed in terms of the primitive flow variable as:
	
	\begin{eqnarray}
		\tau_{ij} = \mu\left[ \left(\frac{\partial u_i}{\partial x_j} + \frac{\partial u_j}{\partial x_i}\right) - \frac{2}{3} \frac{\partial u_k}{\partial x_k}\delta_{ij}\right],
		\label{E4}
		\\
		q_j = -\kappa \frac{\partial T}{\partial x_j} .
		\label{E5}
	\end{eqnarray}
	
Eq. (\ref{E4}) represents the viscous shear stress for the Newtonian fluids, and Eq. (\ref{E5}) is Fourier's law for heat conduction, which relates the heat transfer rate to the temperature gradient in the flow. Note $\delta_{ij}$ is the Kronecker delta, which is equal to 1 when $i=j$ or 0 otherwise. $\mu$ and $\kappa$ are the properties of fluid representing the dynamic viscosity and thermal conductivity, respectively. The dynamic viscosity is either a constant (set through the Reynolds number under investigation) or it is evaluated based on Sutherland's formula for the high-speed flows considered in this investigation. The thermal conductivity is related to viscosity through the Prandtl number ($Pr$) and the heat capacity at constant pressure ($C_p$). This system of equations is closed using the ideal gas assumptions. Note that the Euler equations can be derived by dropping out the viscous terms in the above-mentioned governing equations. They are mentioned here for ease of understanding the development of the solver. 

\begin{eqnarray}
	\frac{\partial \rho}{\partial t}+\frac{\partial (\rho u_j)}{\partial x_j} = 0 \\
	\frac{\partial (\rho u_i)}{\partial t}+\frac{\partial (\rho u_i u_j)}{\partial x_j}  +\frac{\partial p}{\partial x_i} =0 \\
	\frac{\partial (\rho E)}{\partial t}+\frac{\partial (\rho E u_j)}{\partial x_j}+\frac{\partial p u_j}{\partial x_j} = 0
\end{eqnarray}
Note that we conveniently change the Einstein notation of the governing equation to other suitable forms while explaining the convective flux discretisation. Also, we apply the standard second-order central scheme for the Navier-Stokes equation problem. 

	
\subsection{Shock capturing schemes}

This study employs various convective flux discretisation schemes based on shock capturing methods such as Kurganov Noelle Petrova (KNP), Kurganov Tadmor (KT) (these schemes are employed in the density-based default solver in the OpenFOAM package), Hartex Lax Van Leer Contact (HLLC), HLLC scheme with Low Mach number correction (HLLC-LM), HLLC with artificial dissipation (HLLCP), Advection Upstream Splitting Method for all Mach numbers (AUSM+up), and Low Diffusion Flux Splitting Scheme (LDFSS). Their mathematical formulation is outlined here strictly for ease of explanation and convenience.   

This investigation relies on the collocated cell-centred discretisation method; therefore, values from the cell centres to the face centres of the finite control volume are facilitated using the Total Variation Diminishing (TVD) based reconstruction method. Van Leer or Minmod is used in this investigation and is explicitly mentioned for different test cases under investigation.

\subsubsection{rhoCentralFoam}
\textit{rhoCentralFoam} is the default, density-based solver, which is available in the OpenFOAM framework. This scheme uses two simple schemes, Kurganov, Noelle, Petrova (KNP), and Kurganov Tadmor scheme for the convective flux discretization, which is selected by setting the corresponding input flag. The formulation of these schemes is readily available in Ref. \cite{greenshields2010implementation}, and they are discussed here briefly. The same notations are used here to avoid any confusion to the reader. Any divergence term in the governing equation can be evaluated using the finite volume approach by simply integrating over any arbitrary control volume and applying the Gauss divergence theorem, e.g,

\begin{equation}
\int_V \nabla \cdot [\vec{u} \, \Psi] \, \mathrm{d}V 
= \int_S \vec{n} \cdot [\vec{u} \, \Psi] \, \mathrm{d}S 
\approx \sum_f \vec{S}_f \cdot \vec{u}_f \, \Psi_f 
= \sum_f \phi_f \, \Psi_f
\end{equation}
$\vec{u}$ is the velocity vector and $\vec{n}$ is the unit normal vector pointing outwards from the surface. The application of the Gauss divergence theorem results in the surface integral that can be approximated as the summation of volumetric fluxes $\phi_f$, evaluated over the surface of a control volume. Note that $\psi$ is any arbitrary tensor quantity being convected. The straightforward application of KT and KNP schemes lies in the appropriate treatment of the $\phi_f$ volumetric flux and interpolating at the face centres. This is further evaluated as,

\begin{equation}
\sum_f \phi_f \, \Psi_f = \sum_f \left[ \alpha \phi_{f^+} \, \Psi_{f^+} 
+ (1 - \alpha) \phi_{f^-} \, \Psi_{f^-} 
+ \omega \left( \Psi_{f^-} - \Psi_{f^+} \right) \right]
\end{equation}

\begin{figure}[t]
\centering
\includegraphics [width=0.3\textwidth]{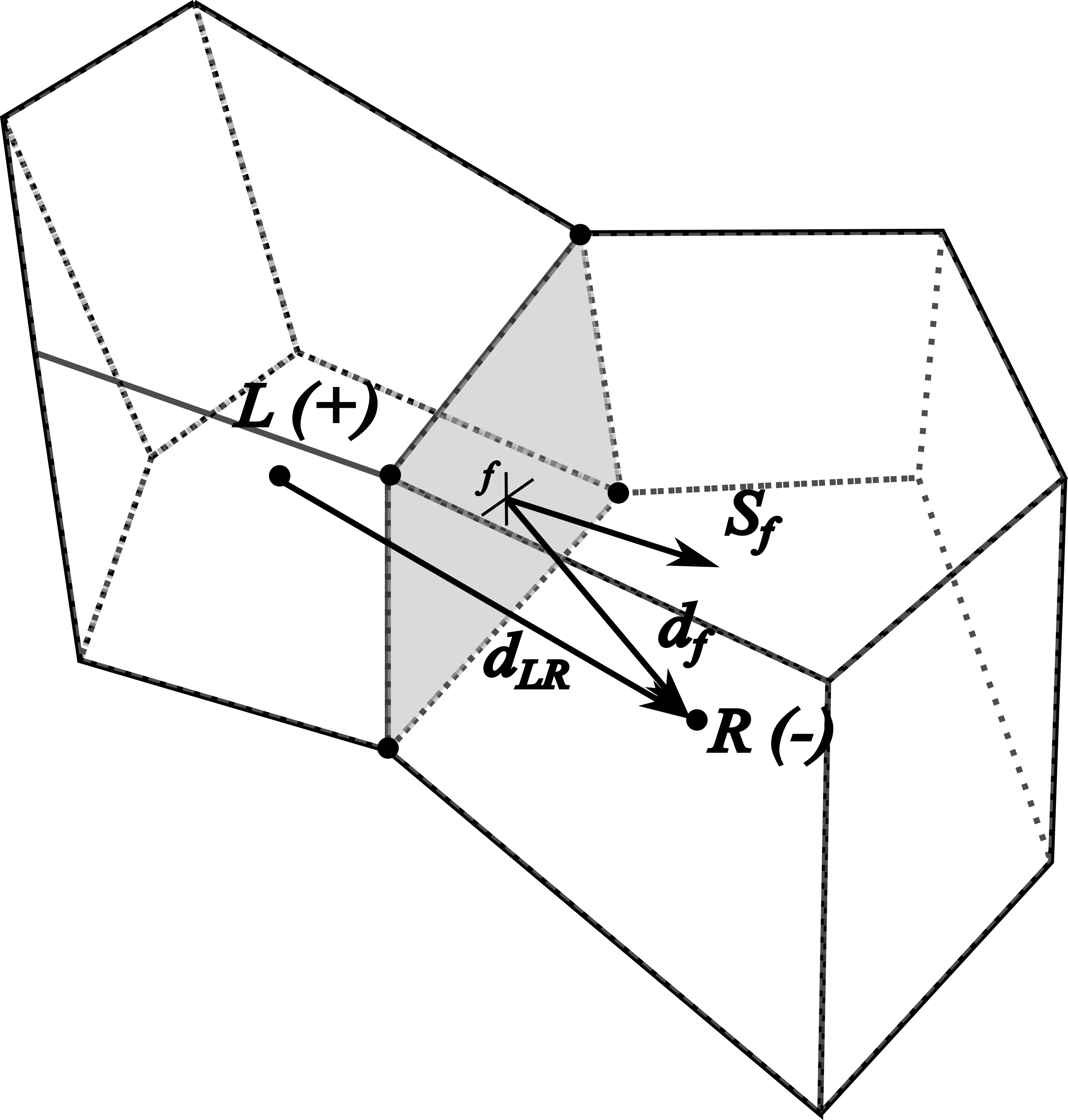}
\caption{Arbitrary control volumes in a grid with the relevant geometrical information. $\vec{S_f}$ is the face area vector perpendicular to the face f, $\vec{d_{LR}}$ is the vector connecting the cell centers L and R, and $\vec{d_f}$ is the vector connecting the face center f and cell centre R}\label{FigureA}
\end{figure}

The first two terms correspond to the segregation of the fluxes based on the unit normal direction vector of the sharing face of the control volumes, with '+' representing the contribution from the interpolation facilitated in the normal direction, and likewise '-' is interpolation facilitated in the reverse direction (see Figure \ref{FigureA} for the clear representation). Some codes also refer to this notation as 'owner' and 'neighbour' respectively. The third term is necessary for those variables that are part of the total derivative in the governing equation. It acts as an extra dissipation to these terms when a discontinuity is detected at the interface. The weighing factor $\alpha$ and the dissipation coefficient $\omega$ are dependent on the schemes being used, i.e., KT or KNP; they are calculated as,

\begin{equation}
\alpha =
\begin{cases}
\frac{1}{2} & \text{for the KT scheme} \\
\frac{\xi_{f^+}}{\xi_{f^+} + \xi_{f^-}} & \text{for the KNP scheme}
\end{cases}
\end{equation}

\begin{eqnarray}
\xi_{f^+} = \max\left(c_{f^+}|\vec{S}_f| + \phi_{f^+},\; c_{f^-}|\vec{S}_f| + \phi_{f^-},\; 0 \right) \\
\xi_{f^-} = \max\left(c_{f^+}|\vec{S}_f| - \phi_{f^+},\; c_{f^-} |\vec{S}_f| - \phi_{f^-},\; 0 \right)
\end{eqnarray}
with \( c_{f^\pm} = \sqrt{\gamma R T_{f^\pm}} \) are the speeds of sound of the gas interpolated onto the interface from the owner and neighbour cells.

The dissipation coefficient is expressed as,
\begin{equation}
\omega_f =
\begin{cases}
\alpha \max\left( \xi_{f^+}, \xi_{f^-} \right) & \text{for the KT scheme} \\
\alpha (1 - \alpha)(\xi_{f^+} + \xi_{f^-}) & \text{for the KNP scheme}
\end{cases}
\end{equation}
The gradient term is handled separately in these schemes. Any gradient terms in the governing equations can be evaluated using the same fundamental application of the Gauss divergence theorem as expressed below:

\begin{equation}
\int_V \nabla \Psi \, \mathrm{d}V = \int_S \vec{n} \Psi \, \mathrm{d}S 
\approx \sum_f \vec{S}_f \Psi_f
\end{equation}
 The application of the KT/KNP scheme results in:
\begin{equation}
\sum_f \vec{S}_f \Psi_f = \sum_f \left[ \alpha \vec{S}_f \Psi_{f^+} + (1 - \alpha) \vec{S}_f \Psi_{f^-} \right]
\end{equation}

\begin{figure}[t]
\centering
\includegraphics [width=0.3\textwidth]{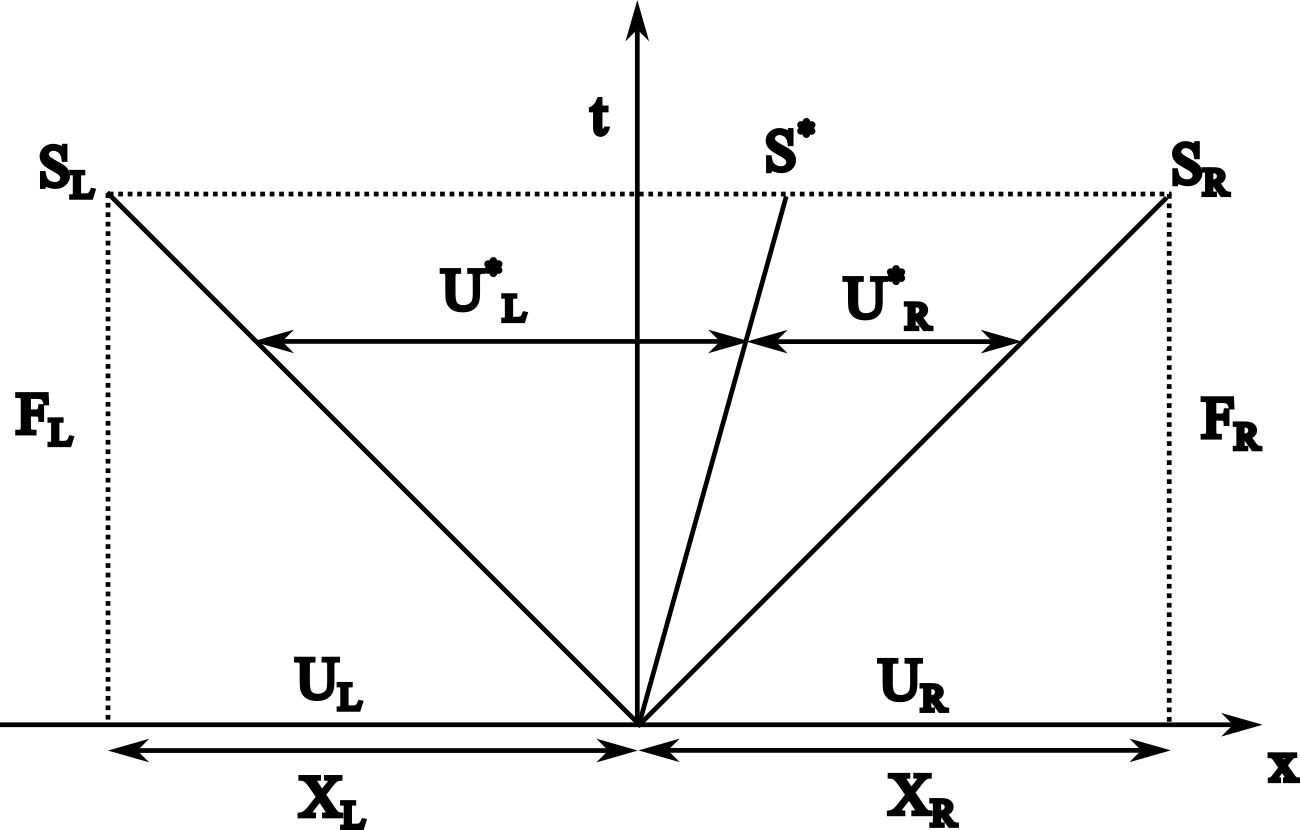}
\caption{A Riemann fan with two intermediate states for the HLLC scheme formulation}\label{riemann_fan}
\end{figure}

\subsubsection{Harten Lax van Leer Contact (HLLC)}
The HLLC scheme is an approximate Riemann solver that builds on the formulation of the HLL scheme \cite{toro1994restoration}. This scheme's features lie in restoring the contact wave in the Riemann problem, which is ignored in the HLL scheme, thus resolving the state of the contact region accurately. For any Riemann problem at an interface, the scheme must determine the state of the flow based on the signal propagation speeds of various discontinuities originating at the Riemann interface, as shown in Figure \ref{riemann_fan}. The state vector and the associated flux function in the Euler equation can be conveniently expressed as.

\begin{equation}
U = 
\begin{bmatrix}
\rho \\
\rho u \\
\rho v \\
\rho w \\
e
\end{bmatrix},
\quad
F = 
\begin{bmatrix}
\rho q \\
\rho u q + p n_x \\
\rho v q + p n_y \\
\rho w q + p n_z \\
(e + p) q
\end{bmatrix},
\label{E6}
\end{equation}  

where \( \rho, u, v, w, e, \) and \( p \) denote density, Cartesian velocity components, total energy per unit volume, and pressure, respectively.  
The directed velocity (or contravarient velocity), \( q = u n_x + v n_y + w n_z \), is the resolution of the velocity component in the \( \vec{n} \) direction, with \( \vec{n} \) as the unit vector, suggesting the orientation of an interface that separates the two distinct states defining the Riemann problem depicted in Figure \ref{riemann_fan}.


The HLLC scheme selects the appropriate state variables and the associated flux functions based on the signal's speed calculation emanating from the interface. The state vector can be selected based on the conditions shown below.,

\begin{equation}
U_{\text{HLLC}} =
\begin{cases}
U_L, & \text{if } S_L > 0, \\
U_L^*, & \text{if } S_L \leq 0 < S^*, \\
U_R^*, & \text{if } S^* \leq 0 \leq S_R, \\
U_R, & \text{if } S_R < 0.
\end{cases}
\label{E7}
\end{equation}

with the interface flux, represented as \( F_{\text{HLLC}} \) which is mathematically written as,

\begin{equation}
F_{\text{HLLC}} =
\begin{cases}
F_L, & \text{if } S_L > 0, \\
F_L^*, & \text{if } S_L \leq 0 < S^*, \\
F_R^*, & \text{if } S^* \leq 0 \leq S_R, \\
F_R, & \text{if } S_R < 0.
\end{cases}
\label{E8}
\end{equation}
The L/R states on the interface can be simply used by using the sharing cell centre values interpolated on the face centre of a cell (using either an upwind scheme for first order accuracy or high-order interpolation schemes); however, the calculation of the state vectors and fluxes in the subsonic region separated by a contact wave is quite involved. The flux function in the intermediate contact region is mathematically expressed as,

\begin{equation}
\begin{aligned}
& F^*_L \equiv F(U^*_L) =
\begin{bmatrix}
\rho_L^* S^* \\
(\rho u)^*_L S^* + p^* n_x \\
(\rho v)^*_L S^* + p^* n_y \\
(\rho w)^*_L S^* + p^* n_z \\
(e^*_L + p^*) S^*
\end{bmatrix} \\
& F^*_R \equiv F(U^*_R) =
\begin{bmatrix}
\rho_R^* S^* \\
(\rho u)^*_R S^* + p^* n_x \\
(\rho v)^*_R S^* + p^* n_y \\
(\rho w)^*_R S^* + p^* n_z \\
(e^*_R + p^*) S^*
\end{bmatrix}
\end{aligned}
\end{equation}

with the intermediate state variables as,

\begin{equation}
U_L^* = 
\begin{bmatrix}
\rho_L^* \\
(\rho u)^*_L \\
(\rho v)^*_L \\
(\rho w)^*_L \\
e^*_L
\end{bmatrix}
=
\Omega_L\begin{bmatrix}
\rho_L(S_L - q_L) \\
(S_L - q_L)(\rho u)_L + (p^* - p_L)n_x \\
(S_L - q_L)(\rho v)_L + (p^* - p_L)n_y \\
(S_L - q_L)(\rho w)_L + (p^* - p_L)n_z \\
(S_L - q_L)e_L - p_L q_L + p^* S^*
\end{bmatrix}
\end{equation}

\begin{equation}
U_R^* = 
\begin{bmatrix}
\rho_R^* \\
(\rho u)^*_R \\
(\rho v)^*_R \\
(\rho w)^*_R \\
e^*_R
\end{bmatrix}
=
\Omega_R\begin{bmatrix}
\rho_R(S_R - q_R) \\
(S_R - q_R)(\rho u)_R + (p^* - p_R)n_x \\
(S_R - q_R)(\rho v)_R + (p^* - p_R)n_y \\
(S_R - q_R)(\rho w)_R + (p^* - p_R)n_z \\
(S_R - q_R)e_R - p_R q_R + p^* S^*
\end{bmatrix}
\end{equation}
with, 
\begin{equation}
\Omega_L \equiv (S_L - S^*)^{-1}, \quad
\Omega_R \equiv (S_R - S^*)^{-1}
\end{equation}
the pressure in the star regions can be conveniently written as,

\begin{equation}
p^* = \rho_L(q_L - S_L)(q_l - S^*) + p_L 
= \rho_R(q_R - S_R)(q_R - S^*) + p_R
\end{equation}
and signal speeds are evaluated as,

\begin{eqnarray} 
S^* = \frac{ \rho_R q_R (S_R - q_R) - \rho_L q_L (S_L - q_l) + p_L - p_R }
{ \rho_R(S_R - q_R) - \rho_L(S_L - q_L) } 
\label{E300001}\
\\
S_L = \min[q_L - c_L, \tilde{q} - \tilde{c}] 
\label{E301}
\\
S_R = \max[q_R + c_R, \tilde{q} + \tilde{c}]
\label{E302}
\end{eqnarray}
$\tilde{(\cdot)}$ are the Roe averages.

\begin{eqnarray}
\tilde{q} = \tilde{u}n_x + \tilde{v}n_y + \tilde{w}n_z, \\
\tilde{u} = \frac{u_L + u_R R_\rho}{1 + R_\rho}  \\
\tilde{v} = \frac{v_L + v_R R_\rho}{1 + R_\rho} \\
\tilde{w} = \frac{w_L + w_R R_\rho}{1 + R_\rho} \\
\tilde{c} = \frac{c_L + c_R R_\rho}{1 + R_\rho} \\
R_\rho = \sqrt{\rho_R / \rho_L},
\end{eqnarray}
This formulation of the HLLC scheme is referred to as HLLC1 in this paper.

\subsubsection{HLLC-LM}

This scheme is suggested by Fleischmann et al. \cite{fleischmann2020shock} cleverly reuses the definition of the HLLC flux schemes for the intermediate state by applying the traditional Rankine-Hugoniot jump conditions across the Riemann fan from the left and right sides of the contact discontinuity and arriving at the same state and averaging them. The jump condition applied across the left state results in this flux function:

\begin{equation}
F_L^*= F_L + S_L \left(U_L^* - U_L \right) \\
\end{equation}
while the jump condition from the other side of the Riemann fan results in,
\begin{equation}
F_L^* = F_R + S_R \left( U_R^* - U_R \right) + S^* \left( U_{L}^* - U_{R}^* \right) 
\end{equation}

They can be simply averaged to arrive at the central formulation of the HLLC scheme,

\begin{equation}
\begin{split}
F_{L}^* &= \frac{1}{2} \left( F_L + F_R \right) + \frac{1}{2} \left[ S_L \left( U_L^* - U_L \right) + S^* \left( U_L^* - U_R^* \right) \right]\\
&\qquad + S_R \left( U_R^* - U_R \right) 
\end{split}
\end{equation}

Similarly, the modified expression of $F_R^*$ can also be obtained in a similar fashion, expressed as,

\begin{equation}
\begin{split}
F_{R}^* &= \frac{1}{2} \left( F_L + F_R \right) + \frac{1}{2} \left[ S_L \left( U_L^* - U_L \right) - S^* \left( U_L^* - U_R^* \right) \right]\\
&\qquad + S_R \left( U_R^* - U_R \right) 
\end{split}
\end{equation}

These alternative flux functions written in the central form are referred to as HLLC2 in this paper. The authors identified that the $S^*$ is the acoustic dissipation, which is over-compensated compared to the convective diffusion coefficients $S_L$ and $S_R$ under the low Mach number limit, where the normal components to the grid-aligned discontinuity diminish. Therefore, they aimed to fine-tune these parameters such that diffusion coefficients are balanced under the low Mach number limit, alternatively, where the strong grid-aligned discontinuity exists. They simply regulated the convective diffusion coefficient by this simple modification to the central form of the HLLC scheme:

\begin{align}
S_L^{HLLC-LM} &= \phi \cdot S_L, \quad 
S_R^{HLLC-LM} = \phi \cdot S_R 
\end{align}

with

\begin{align}
\phi = \sin\left( \min\left( 1, \frac{Ma_{local}}{Ma_{limit}} \right) \cdot \frac{\pi}{2} \right) 
\end{align}

and

\begin{align}
Ma_{local} = \max\left( \left| \frac{q_L}{c_L} \right|, \left| \frac{q_R}{c_R} \right| \right) 
\end{align}

The $\phi$ function returns unity when the threshold Mach number $Ma_{limit}$ is exceeded. This study uses $Ma_{limit} = 0.1$ for all the cases used in this investigation \cite{fleischmann2020shock}.

\subsubsection{HLLCP}
We adopt another variation of the HLLC scheme correction, whose principle lies in dissipating the erroneous flux calculation behind the grid-aligned discontinuity number region by applying extra dissipation in that region as suggested in Ref. \cite{xie2019accurate}. According to the authors of this reference, the modified HLLC scheme can be expressed as:

\begin{equation}
F_{\text{HLLC-P}} = F_{\text{HLLC}} + \Phi_p 
\end{equation}
where $\phi_p$ is the extra added dissipation. For a two-dimensional flow, it is expressed as,

\begin{equation}
\Phi_p = (f - 1) \frac{S_L S_R}{S_R - S_L} \cdot \frac{1}{1 + |\widehat{M}|} \cdot \frac{\Delta p}{\widehat{c}^2}
\begin{pmatrix}
1 \\
\widehat{u} \\
\widehat{v} \\
\frac{1}{2} \widehat{q}^2
\end{pmatrix}.
\end{equation}

In this equation, $S_L$ and $S_R$ are defined in Eqs. \ref{E301} and \ref{E302}, $\Delta p$ determines the pressure difference across the interface, along with  
$\widehat{M}, \widehat{c}, \widehat{u}, \widehat{v},$ and $\widehat{q}$ denote Roe-averaged Mach number, acoustic speed, cartesian velocity components, and the directed (contravariant or normal) velocity computed at the interface, respectively. The $f$ is a sensor that detects smooth and non-smooth regions in the flow field, determines the pressure weight, which is calculated as shown:

\begin{equation}
\begin{split}
f &= \min(f_{p,i,j+1/2}, f_{p,i-1/2,j}, f_{p,i-1/2,j+1}, f_{p,i+1/2,j}, \\
&\qquad  f_{p,i+1/2,j+1})^3    
\end{split}
\end{equation}

This function ensures that the extra added dissipation is diminished in the smooth flow region as $f$ approaches unity. They also suggested a low Mach number treatment for the all Mach version of their robust scheme by modifying the pressure term in the contact region. They expressed this all Mach number version of the flux in the following way:

\begin{equation}
\Phi^*_{\text{AM--HLLC--P,K}} =
\begin{pmatrix}
\rho_K^* S^* \\
\rho_K^* S^* \left[ u_K + n_x (S^* - q_K) \right] + p^{***} n_x \\
\rho_K^* S^* \left[ v_K + n_y (S^* - q_K) \right] + p^{***} n_y \\
S^* \left( \rho_K^* e_K^* + p^* \right)
\end{pmatrix}_{ij}
+ \Phi_p,
\end{equation}
\\
with $K = L/R$ and $(\cdot)_{ij}$ are Roe averaged values. Further $p^{***}$, which is a pressure correction term, expressed as:

\begin{eqnarray}
p^{***} = f p^{**} + (1 - f) p^* \\
p^{**} = \theta p^* + (1 - \theta) \frac{p_L + p_R}{2} \\
\theta = min(max(|M_L|, |M_R|),1) \\
M_L = \frac{q_L}{c_L}, M_R = \frac{q_R}{c_R} 
\end{eqnarray}

The blending of $p^{**}$ and $p^{*}$ ensures low Mach number treatment is abolished in the supersonic regime of the flow, with the simultaneous activation of the pressure dissipation term as explained above. 

\subsubsection{Advection Upstream Splitting Method (AUSM+up)}

The AUSM-type family of schemes \cite{liou2006sequel} is built upon a convenient decomposition of the convective flux vector into separate advective and pressure components, i.e.
\begin{eqnarray}
	F = F^{(c)} + F^{(p)} 
\end{eqnarray}

This is further expressed  as

\begin{eqnarray}
	F = \dot{m}\psi + F^{(p)} 
\end{eqnarray}

Here, $\dot{m} = \rho U$ denotes the mass flux, which carries the information about the flow direction, and $\psi = (1, u, v, w, H)^T$ represents the vector of advected quantities. The pressure flux vector is given by $F^{(p)} = (0, pn_x, pn_y, pn_z, 0)^T$, where $\hat{n} = (n_x, n_y, n_z)^T$ is the unit normal vector to an arbitrary surface in the $x$, $y$, and $z$ directions, respectively. The interface flux thus can be written as 

\begin{eqnarray}
	F_{f} = \dot{m}_{f} \psi_{L/R} + F^{(p)}_{f}
\end{eqnarray}
	
In this notation, the subscript 'f' refers to quantities evaluated at the common face of the control volume, whereas $\psi_{L/R}$ denotes the advected quantity (see Figure \ref{FigureA}), which can be computed using a simple upwind scheme as illustrated here

\begin{eqnarray}
	\psi_{L/R} = 
	\begin{cases}
		\psi_{L} &\textit{if $\dot{m}_{f} >$ 0}\\
		\psi_{R} &\textit{otherwise} 
	\end{cases}
\end{eqnarray}

Consequently, the mass flux at the interface, $\dot{m}_{f}$, is also determined using the same strategy. 

\begin{eqnarray}
	\dot{m}_{f} = a_{f}M_{f} 
	\begin{cases}
		\rho_{L} &\textit{if $M_{f} >$ 0}\\
		\rho_{R} &\textit{otherwise} 
	\end{cases}    
\end{eqnarray}
	
The primary goal is to evaluate the interface mass flux $\dot{m}_{f}$ and the interface Mach number $M_{f}$ to fully define the flux function across the interface. The expression for $M_{f}$ can be directly written as:

\begin{eqnarray}
	M_{f} = \mathscr{M}_{(m)}^{+}(M_L) + \mathscr{M}_{(m)}^{-}(M_R) + M_p
\end{eqnarray}

Here, 
\begin{eqnarray}
	M_{L/R} = \frac{u_{L/R}}{a_{f}}
\end{eqnarray}

Here, $u = \vec{U} \cdot \hat{n}$, with $\hat{n}$ is the unit normal vector to the cell interface (equivalent to the directed velocity, $q$ used in the HLLC scheme notation. We change this notation to maintain consistency with notation used in Ref. {\cite{liou2006sequel}}), and $a_{f}$ represents either the arithmetic or geometric mean of $a_L$ and $a_R$, which are the speeds of sound in the left and right cells, respectively. The terms $\mathscr{M}_{(m)}^{\pm}$ are interpolation polynomials of order $m$ for the left and right sides of the interface. $M_p$ is a diffusion term introduced to suppress spurious oscillations near discontinuities, and it is related to the pressure difference across the cell face. These definitions are provided below.

\begin{eqnarray}
	\mathscr{M}^{\pm}_{(1)}(M) = \frac{1}{2}(M\pm \lvert M \rvert) \\
	\mathscr{M}^{\pm}_{(2)}(M) = \pm \frac{1}{4}(M\pm 1)^2 \\
	\mathscr{M}^{\pm}_{(4)}(M) = 
	\begin{cases}
		\mathscr{M}_{(1)}^{\pm} &\textit{if $ \lvert M \rvert \geq$ 1}\\
        \\
		\mathscr{M}_{(2)}^{\pm} (1 \mp 16 \beta           \mathscr{M}_{(2)}^{\mp}) &\textit{otherwise} 
	\end{cases}\\
	M_p = -\frac{K_p}{f_a}max(1 - \sigma \Bar{M}^2 , 0)\frac{p_R-p_L}{\rho_{f}a_{f}^2}, \sigma \leq 1
\end{eqnarray}
	
where, $\Bar{M}$ and $\rho_{f}$ is evaluated as,
\begin{eqnarray}
	\Bar{M}^2 = \frac{u_L^2 + u_R^2}{2 a_{f}^2}\\
	\rho_{f} = \frac{\rho_L + \rho_R}{2}
\end{eqnarray}

Here, $f_a$ is introduced in this expression as the scaling factor, which is necessary for the formulation of all Mach number flows. It is evaluated as
\begin{eqnarray}
	f_a = M_0(2 - M_0) 
\end{eqnarray}

where $M_0$ is calculated using the expression,
\begin{eqnarray}
	M_0^2 = min(1, max(\Bar{M}^2, M_{co}^2)) 
\end{eqnarray}
	
with $M_{co}$ as the cut-off Mach number set as the Mach number of the free stream, $M_{\infty}$. On the similar lines, pressure flux, $F^{(p)}_{f}$ is evaluated as,
	
\begin{eqnarray}
	F^{(p)}_{f} = \mathscr{P}_{(n)}^{+}(M_L)p_L + \mathscr{P}_{(n)}^{-}(M_R)p_R + p_u
\end{eqnarray}

in which 'n' is the order of the polynomial ${P}_{(n)}^{\pm}$, which can be 1, 3, or 5. In this investigation, $5^{th}$ order polynomial is used for higher order accuracy. It is expressed as,

\begin{eqnarray}
	\mathscr{P}^{\pm}_{(5)}(M) = 
	\begin{cases}
		\frac{1}{M}\mathscr{M}_{(1)}^{\pm} &\textit{if $ \lvert M \rvert \geq$ 1}\\
        \\
		\mathscr{M}_{(2)}^{\pm} [(\pm 2 - M) \mp 
		16 \alpha M \mathscr{M}_{(2)}^{\mp}] 
		&\textit{otherwise} 
	\end{cases}
\end{eqnarray}
	
with,
\begin{eqnarray}
	\alpha = \frac{3}{16} (-4 + 5 f_a^2) \\
	\beta = \frac{1}{8}
\end{eqnarray}
	
$p_u $ is the pressure dissipation term, which is evaluated as the difference in velocity across the interface and is calculated as,
	
\begin{equation}
	p_u = -K_u \mathscr{P}_{(5)}^{+}(M_L) \mathscr{P}_{(5)}^{-}(M_R) (\rho_L + \rho_R)(f_a a_{f}) (u_R - u_L)
    \\
\end{equation}
\\	
This investigation uses the following values of constants as suggested in Ref.\cite{liou2006sequel}, $K_p = 0.25$, $K_u = 0.75$ \& $\sigma = 1.0$

\subsubsection{Low Diffusion Flux Splitting Scheme (LDFSS)}
The LDFSS scheme \cite{edwards1997low} is another popular scheme in the Flux Vector splitting category that has been proven effective in resolving the complex flow features to a great extent. The LDFSS scheme relies on the traditional splitting of the convection flux into advective and pressure fluxes and treating them individually, similar to the AUSM-type scheme. It can be represented as (same notation used in Ref. \cite{edwards1997low} is used for the sake of convenience),

\begin{equation}
E = E^c + E^p = \rho a M \tilde{E} + p \tilde{E}^p
\end{equation}

where,

\begin{equation}
\tilde{E} = \begin{bmatrix}
1 \\
u \\
v \\
H
\end{bmatrix}, \quad
\tilde{E}^p = \begin{bmatrix}
0 \\
n_x \\
n_y \\
0
\end{bmatrix} 
\end{equation}
 with, 
\begin{eqnarray}
H = \frac{1}{\rho} (E_t + p) = \gamma e + \frac{1}{2} (u^2 + v^2)      
\end{eqnarray}
The contravariant Mach number (or directed Mach number) $M$ is given as (for example, in a two-dimensional flow)
\begin{align}
M = \frac{1}{a} \left[ u n_x +  v n_y \right] 
\end{align}
The advective portion of the convective flux for the LDFSS2 scheme is expressed as,

\begin{equation}
E^c_{f} = \tilde{a}_{f} [\rho_L C^+ \tilde{E}_L + \rho_R C^- \tilde{E}_R]
\end{equation}
with,

\begin{eqnarray}
\tilde{a}_{f} = a_{f} \frac{
\sqrt{(1 - M_{\text{ref}}^2) M_{f}^2 + 4 M_{\text{ref}}^2}
}{
1 + M_{\text{ref}}^2
}
\\
M_{\text{ref}}^2 = \frac{V_{\text{ref}}^2}{a_{f}^2} = \min\left( |\vec{V_{f}}|^2, \kappa V_\infty^2, a_{f}^2 \right)/a_{f}^2
\end{eqnarray}
with $\kappa = 0.25$ and the interface speed of sound $a_{f}$ is evaluated by simple geometric averaging in this study.
\begin{eqnarray}
    a_{f} = \sqrt{a_La_R}
\end{eqnarray}
This treatment of $\tilde{a}_{f}$ (often referred to as numerical speed of sound \cite{edwards1997low}) allows the scheme for enabling the preconditioning in the low Mach number regime. Subsequently, this numerical speed of sound shall be used further to define the Mach numbers wherever appropriate. The functions $C^+$ and $C^-$ for this scheme are evaluated using, 

\begin{eqnarray}
C^+ = C_{VL}^+ - M_{f}^+  \\
C^- = C_{VL}^- + M_{f}^-
\end{eqnarray}
with, $ C_{VL}^+$ and $ C_{VL}^-$ is defined as,

\begin{align}
C_{VL}^+ = \alpha_L^+ (1.0 + \beta_L) M_L - \beta_L M_{L}^+ \\
C_{VL}^- = \alpha_R^- (1.0 + \beta_R) M_R - \beta_R M_{R}^-
\end{align}
and the split Mach number \( M^\pm \) is calculated as,

\begin{equation}
M_{L,R}^\pm = \pm \tfrac{1}{4} (M_{L,R} \pm 1)^2
\end{equation}
whereas, the functions \( \alpha^\pm \) and \( \beta \) ensure appropriate sonic-point transition behavior, defined as, 
\begin{align}
\alpha_{L,R}^\pm &= \tfrac{1}{2} \left[ 1.0 \pm \text{sgn}(M_{L,R}) \right] \\
\beta_{L,R} &= - \max\left[ 0, 1 - \text{int}(|M_{L,R}|) \right]
\end{align}
The modified definition of Mach number using the numerical speed of sound is:

\begin{equation}
M_{L,R} = \frac{1}{\tilde{a}_{f}}
\left[
 u_{L,R} n_x + v_{L,R}n_y
\right]
\end{equation}
In this work, the following definitions for \( M_{f}^{\pm} \) are utilized:

\begin{eqnarray}
M_{f}^{+} = M_{f} \left(1 - \frac{1}{2} \cdot \frac{p_L - p_R}{\rho_L V_{\text{ref}}^2} \right) \\
M_{f}^{-} = M_{f} \left(1 + \frac{1}{2} \cdot \frac{p_L - p_R}{\rho_R V_{\text{ref}}^2} \right)
\end{eqnarray}
and 

\begin{equation}
M_{f} = \frac{1}{4} \beta_L \beta_R \left( \sqrt{\frac{1}{2} (M_L^2 + M_R^2)} - 1.0 \right)^2
\end{equation}
On the other hand, the pressure flux is governed by,

\begin{equation}
\begin{split}
E^p_{1/2} &= \tilde{E_p} \left[\frac{1}{2}(p_L + p_R) 
+ \frac{1}{2}(p_L - p_R)(D_L^{+} - D_R^{-}) \right]\\
 &\quad + \frac{1}{2}  \left[(\rho_L + \rho_R) \, V_{\text{ref}}^2 \left(D_L^{+} + D_R^{-} - 1\right)\right]    
\end{split}
\end{equation}

\( D^\pm \) are calculated as a function of Mach number, and it is conveniently expressed using:

\begin{equation}
D^\pm_{L,R} = \alpha^\pm_{L,R}(1 + \beta_{L,R}) - \beta_{L,R} P^\pm_{L,R}     
\end{equation}
with the subsonic pressure splitting $P^{\pm}_{L,R}$ evaluated as, 

\begin{equation}
P^\pm_{L,R} = \frac{1}{4}(M_{L,R} \pm 1)^2 (2 \mp M_{L,R}) 
\end{equation}

	\subsection{Time integration scheme}


Another distinctive aspect of the implemented solvers is their improved time integration scheme for the governing equations, surpassing the method used in the default \textit{rhoCentralFoam} solver provided in the OpenFOAM package. The standard solver adopts an inviscid-predictor, viscous-corrector strategy \cite{greenshields2010implementation}, where the Euler equations are solved using explicit time integration, followed by an implicit time integration in the corrector step. The inviscid predictor equations for the momentum and energy equations, are shown below for convenience \cite{greenshields2010implementation}: 

\begin{eqnarray}       
       \frac{\partial (\rho u_i)}{\partial t}+\frac{\partial (\rho u_i u_j)}{\partial x_j}  +\frac{\partial p}{\partial x_i} = 0 
		\label{E2000}
		\\
		\frac{\partial (\rho E)}{\partial t}+\frac{\partial (\rho E u_j)}{\partial x_j}+\frac{\partial p u_j}{\partial x_j} = \frac{\partial \tau_{ij} u_i}{\partial x_j}
		\label{E3000}.
\end{eqnarray}
Note that the work done by shear stresses is trivially zero when the flow field is inviscid ($\mu = 0$) while it is calculated explicitly in the predictor step when the full Navier-Stokes equation is solved. The shear stress term is evaluated as shown in Equation \ref{E4}. Similarly, the corresponding viscous predictor equations are written as,

\begin{equation}
\begin{split}
\left[\frac{\partial (\rho u_i)}{\partial t} \right]_V 
&= \frac{\partial}{\partial x_j}\left(\mu \frac{\partial u_i}{\partial x_j} \right) \\
&\quad + \frac{\partial}{\partial x_j}\left\{
\mu\left(\frac{\partial u_j}{\partial x_i} - \frac{2}{3} \frac{\partial u_k}{\partial x_k}\delta_{ij}\right)
\right\}_{\text{explicit}} \\
\left[\frac{\partial (\rho c_v T)}{\partial t} \right]_V 
&= \frac{\partial}{\partial x_j}\left(\kappa \frac{\partial T}{\partial x_j} \right)
\end{split}
\end{equation}


The explicit first-order time integration in the predictor step can lead to numerical instabilities at high CFL numbers. This solver incorporates the Runge-Kutta (RK) time integration method and the different flux scheme outlined to enhance numerical stability. Specifically, it employs a low-storage, four-stage, third-order accurate RK4 \cite{li2020scalability} time integration strategy for the inviscid-predictor step. The temporal integration is expressed as:

\begin{eqnarray}
    U^{(n)} = U^{(0)} - \Delta t \alpha_n \mathscr{R}(U^{n-1}) 
\end{eqnarray}
Here, $U$ denotes the set of integration variables in the governing equations, and $\mathscr{R}(U)$ represents the residual inviscid fluxes, which are evaluated using one of the implemented schemes based on the integration variables. The indices $n$ and $n-1$ refer to the current and previous stages, respectively. This integration method proceeds through four stages, with corresponding stage coefficients given by $\alpha_1 = 1/4$, $\alpha_2 = 1/3$, $\alpha_3 = 1/2$, and $\alpha_4 = 1$. The time step $\Delta t$ is computed based on the CFL stability criterion. The robust RK4 integration approach permits a CFL number of up to 0.5 without introducing spurious numerical oscillations in the solution, in contrast to the standard \textit{rhoCentralFoam} solver. For the viscous predictor step, a simple first-order implicit Euler integration can be applied. A detailed description of the solver algorithm is provided in Section \ref{algo}.

 \subsection{Solution algorithm \label{algo}}
As stated earlier, the sequential temporal integration of the Euler equation creates a lag in the updated quantities, specifically between the velocity and temperature fields, thus can generate undesirable numerical noise. The novel integration of the RK4 scheme in the predictor-corrector step helps alleviate this issue. The detailed algorithm is outlined in Algorithm 1.

\newtcolorbox[auto counter, number within=section]{algobox}[1][]{%
  enhanced,
  breakable,
  colback=white,
  colframe=black,
  fonttitle=\bfseries,
  coltitle=white,
  boxrule=0.3pt,
  arc=0.3mm,
  left=0.8mm,
  right=0.8mm,
  top=0.5mm,
  bottom=0.5mm,
  fontupper=\footnotesize,
  title={Algorithm 1: Integration of RK4 scheme into the predictor-corrector algorithm },
  #1
}

\begin{algobox} [label={alg:rk4_predictor}]

1: Read initial conditions and inputs for the constants (gas properties), CFL, and initial $\Delta t$ \\
2: \textbf{while} $t \leq t_{\text{end}}$ \textbf{do} \hfill \textit{ Start time loop for unsteady integration} \\
3: Save $\rho$, $(\rho \vec{u})$, $(\rho E)$ from previous time step (or initial condtions) in $\rho_{\text{old}}$, $(\rho \vec{u})_{\text{old}}$, $(\rho E)_{\text{old}}$ \\
4: Set $j = 1$, $N = 4$, $\alpha_1 = 0.25$, $\alpha_2 = 1/3$, $\alpha_3 = 0.5$, $\alpha_4 = 1.0$ \\
5: \textbf{while} $j \leq N$ \textbf{do} \hfill \textit{ Start RK4 sub-stage integration} \\
6: Evaluate $\rho_f^j$, $(\rho \vec{u})_f^j$, $(\rho E)_f^j$ from cell centres onto face centres using any reconstruction scheme \\
7: Compute: $\vec{u}_f^j = (\rho \vec{u})_f^j / \rho_f^j$, $p_f^j = \rho_f^j R T_f^j$, $a_f^j = \sqrt{\gamma R T_f^j}$ \\
8: Compute convective fluxes (mass, momentum, energy) using the implemented scheme (e.g., HLLC) and store in \texttt{rhoFlux}, \texttt{momFlux}, \texttt{enFlux} \\
9: Solve density equation: $\rho = \rho_{\text{old}} + \alpha_j \Delta t (-\texttt{rhoFlux})$ \\
10: Solve momentum predictor: $(\rho \vec{u}) = (\rho \vec{u})_{\text{old}} + \alpha_j \Delta t (-\texttt{momFlux})$ \\
11: Compute velocity field $\vec{u} = \rho \vec{u}/\rho$ \\
12: Update boundary conditions on $\vec{u}$ \\
13: Update boundary field of $(\rho \vec{u})$ \\
14: \textbf{if} $\mu \neq 0$ \textbf{and} $j==4$ \textbf{then} Compute $\tau$, solve momentum corrector equation (implicit) \\
15: Update BCs of $\vec{u}$ and boundary field of $(\rho \vec{u})$ \\
16: \textbf{end if} \\
17: Solve energy predictor: $\rho E = (\rho E)_{\text{old}} + \alpha_j \Delta t (-\texttt{enFlux})$ \\
18: Evaluate internal energy: $e = \rho E/\rho - 0.5|\vec{u}|^2$ \\
19: Find temperature from $e = C_v(T) T$ using Newton-Raphson \\
20: Update BCs on $T$ and update field data of $e$ and $\rho E$ \\
21: \textbf{if} $\mu \neq 0$ \textbf{and} $j==4$ \textbf{then} Compute $\tau_U$, solve energy corrector equation (implicit) \\
22: Calculate temperature and update BCs of $T$ and boundary field of $\rho E$ \\
23: \textbf{end if} \\
24: Compute pressure field: $p = \rho R T$ \hfill \textit{ For ideal gas} \\
25: Update pressure BCs and boundary field of $\rho$ using $\rho = p / RT$ \\
26: \textbf{end while} \hfill \textit{ End RK4 sub-stage} \\
27: $t = t + \Delta t$, with $\Delta t$ from CFL condition \\
28: \textbf{end while} \hfill \textit{ End time loop}
\end{algobox}

\section{Results: Solution of Euler equations \label{Section3}} 

\subsection{One-dimensional tests}
We first initiate a discussion on the solvers' behaviour in terms of robustness, accuracy, and their dispersion characteristics by simulating simple one-dimensional Riemann problems but using the Euler equations. The absence of real physical viscosity for these governing equations enables us to judge the inherent numerical diffusion/dispersion associated with these shock-capturing schemes. We first demonstrate the modified shock tube problem, which contains a criticality in terms of admission of non-physical expansion shock waves if they don't satisfy the entropy positivity, and further simulate these Riemann problems with increasing complexities with finer mesh sizes, which ultimately reveals the dispersive behaviour of the schemes (if any). 

Note, we operate the default solver with Euler time integration for the temporal terms in the governing equations to lay the benchmark for the default implemented schemes in the OpenFOAM framework. The cut-off Mach number for the HLLC-LM scheme is 0.1 for all the cases unless otherwise stated. Further, we use the non-dimensionalisation procedure outlined in Ref. \cite{woodward1984numerical,colella1984piecewise} to represent the properties of the working medium. 

\subsubsection{Modified Sod's shock tube}


An Initial Boundary Value Problem (IBVP), in which the initial conditions feature a discontinuity at a certain place in the domain, can be used to simulate a shock tube kind of problem. We use Toro's modified Sod's Shock Tube \cite{toro2013riemann} for this numerical test, which has a left traveling sonic expansion wave, a right propagating shock wave, and a contact discontinuity. The eigenvalues of the features change signs at a particular point in the expansion wave, which can make it difficult for flux schemes to reproduce this behavior. An unwanted expansion shock feature that defies the entropy requirement and is frequently referred to as an \textit{entropy glitch} is produced when these numbers cannot be replicated. The initial discontinuity, or $x_0=0.5$, is positioned at the center of the 400 finite volume cells that make up the computational spatial domain, which extends from [0 1]. Below are the fluid states $(\rho, u, p)$ on the left and right of the diaphragm as an initial condition. At the domain's endpoints, transmissive criteria are utilized. Regardless of the strategies employed in this investigation, the CFL number is kept constant at 0.2 for this numerical test.

\begin{figure*}[t]
\centering
\includegraphics [width=0.6\textwidth]{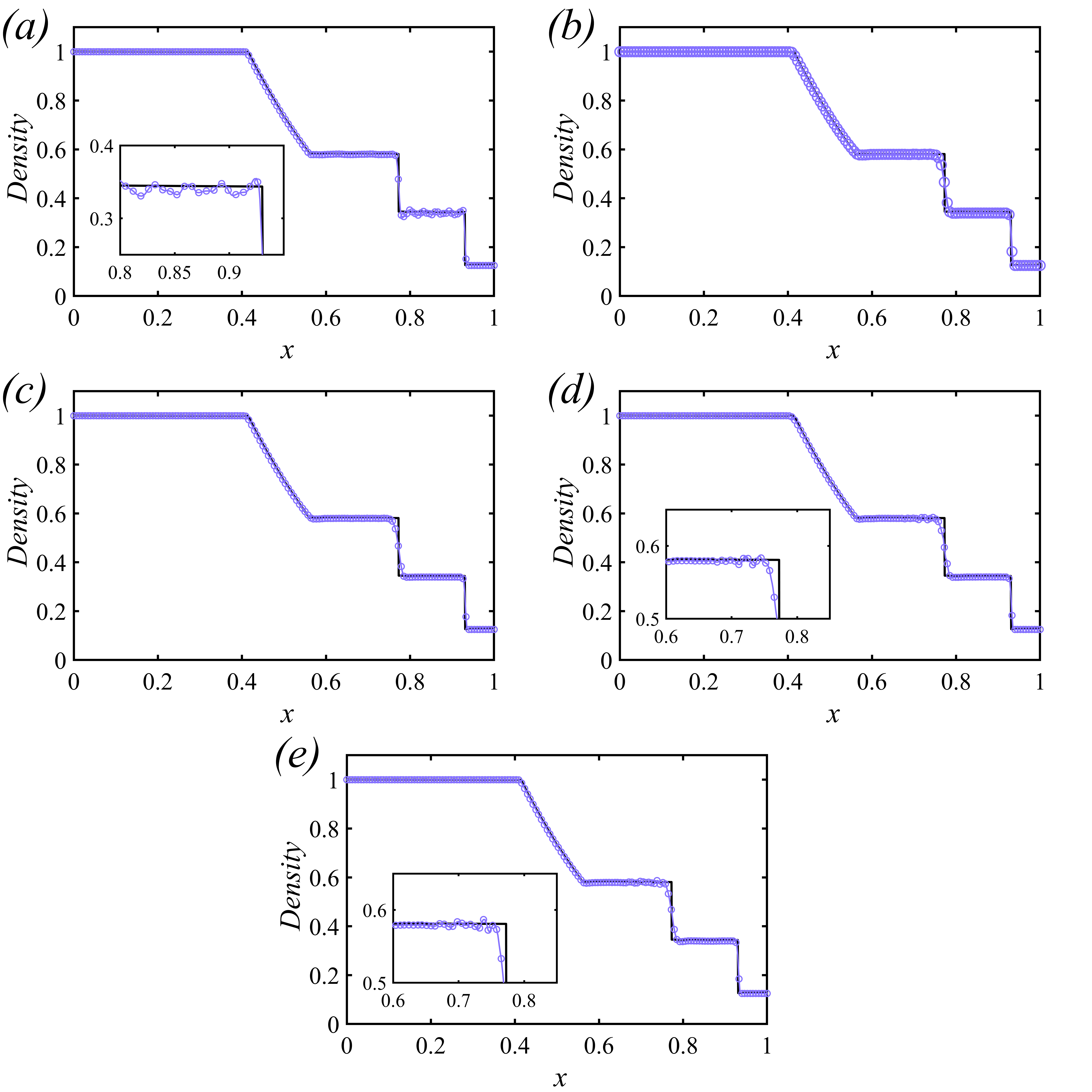}
\caption{Spatial density distribution at $t_f = 0.2$ for the modified shock tube problem with different flux schemes. The solid black line represents the analytical solution \cite{toro2013riemann}, and the blue circles represent the numerical solution. (a) \textit{rhoCentralFoam} (b) HLLC-LM (c) HLLCP (d) AUSM+up (e) LDFSS}\label{Figure1}
\end{figure*}

\[
    (\rho, u, p) =
    \begin{cases}
	(1, 0.75, 1) & \text{if } x \leq 0.5, \\
	(0.125, 0, 0.1) & \text{if } x > 0.5.\\
    \end{cases}
\]
	

Figure \ref{Figure1} represents the spatial density distribution of density at $t_f = 0.2$. The default solver \textit{rhoCentralFoam} in Figure \ref{Figure1} (a) generates a spurious oscillation behind the right propagating shock wave and is a non-monotone solution (also see the inset of this Figure). As a result, even with a low CFL number, the default solver fails to produce a stable solution and may eventually cause spurious oscillations. The contact and expansion waves (including head and tip) and other characteristics of the solution derived from this scheme precisely overlap with the analytical solution.

Similar observations are derived for the HLLC family schemes in Figure \ref{Figure1} (b)-(e), except that they produce a monotone solution free from any spurious oscillations. On the contrary, these numerical instabilities reappear in the contact wave region for the flux vector splitting type schemes due to an imbalance between the split pressure terms that appear in their formulations \cite{moguen2020diffusion}. However, they are very feeble compared to the scales of fluctuations present in the default solver. It is worth mentioning that all of these schemes produce correct wave speeds; therefore, the average location of the features mentioned above is in good agreement with the theory. On the positive side, none of these schemes violated the entropy condition that can form an expansion shock at the sonic point.

\begin{figure}[t]
\includegraphics [width=0.5\textwidth]{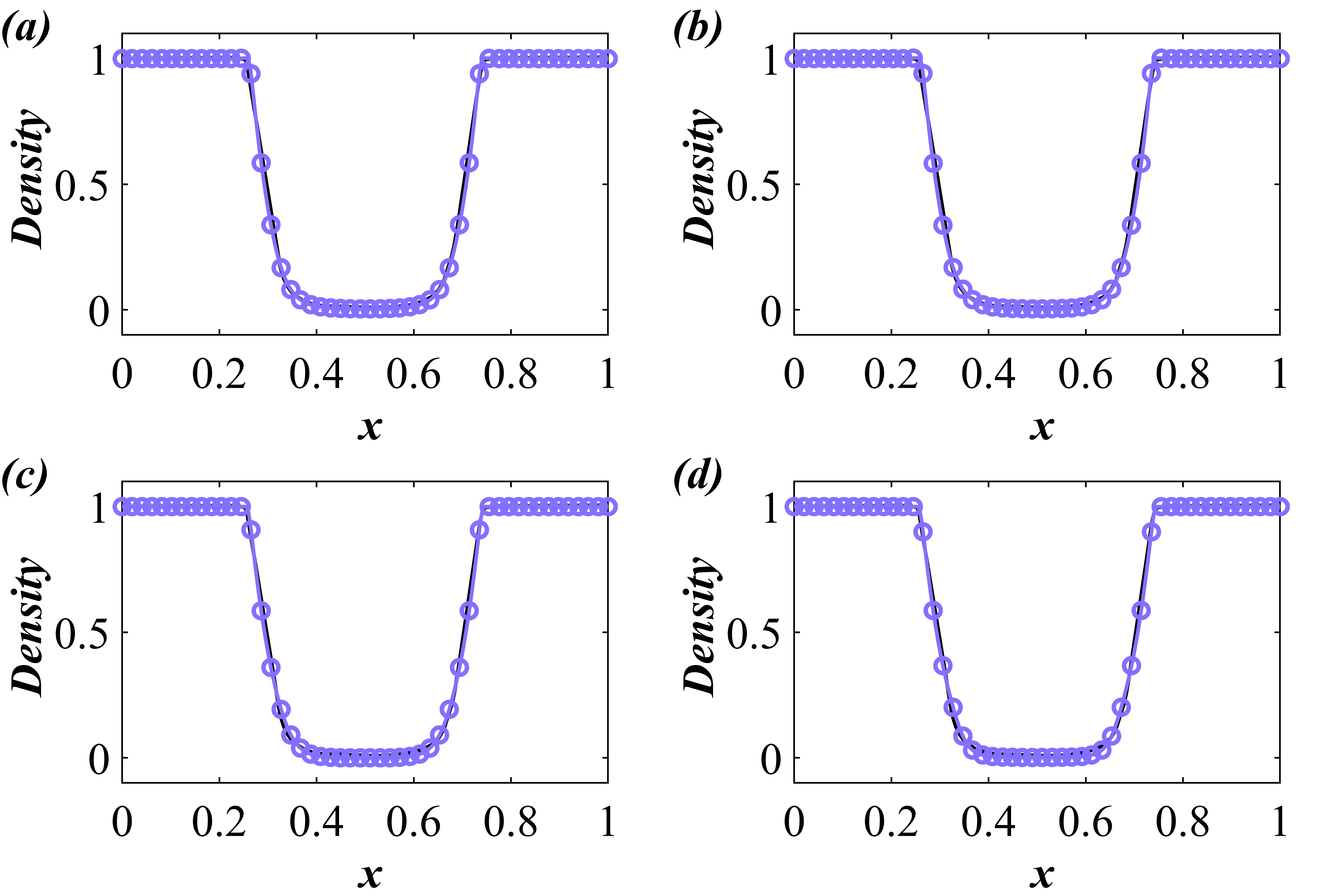}
\caption{Spatial density distribution at $t_f = 0.1$ for double rarefaction problem with different flux schemes. The solid black line represents the analytical solution \cite{hu2013positivity}, and the blue circles represent the numerical solution. (a) HLLC-LM (b) HLLCP (c) AUSM+up (d) LDFSS}\label{Figure2}
\end{figure}

\subsubsection{Double rarefaction waves}
The second problem in this series tests the robustness of the employed solvers to accurately resolve the two expansion waves moving in opposite directions, creating a contact region at near-zero pressure (vacuum) in the initial discontinuity location. This test case is ideal to test the scheme's robustness for the rarefied flow \cite{toro2013riemann} and to check its positivity preserving characteristics \cite{hu2013positivity}. The domain spans [0 1], discretised with 400 cells and simulated till the final time, $t_f = 0.1$. The initial conditions are listed below \cite{phongthanapanich2024shock}.

\[
    (\rho, u, p) =
    \begin{cases}
	(1, -2, 0.1) & \text{if } x \leq 0.5, \\
	(1, 2, 0.1) & \text{if } x > 0.5.\\
    \end{cases}
\]

The density distribution is illustrated in Figure \ref{Figure2} for this test case. The schemes HLLC-LM (\ref{Figure2} (a)), HLLCP (\ref{Figure2} (b)), AUSM+up (\ref{Figure2} (c)), and LDFSS (\ref{Figure2} (d)) produce stable results matching accurately with the exact solution; however, the default solver \textit{rhoCentralFoam} fails to provide any solution thus failing to ensure positivity preserving property. We monitored the temperature evolution (not shown) with iterations of the solver and found that it remains unbounded, producing negative temperatures, which is unphysical; However, the other schemes with RK4 time integration omit this behaviour and generate a stable solution as shown above.   

\subsubsection{Shu and Osher problem}

\begin{figure}[t]
\centering
\includegraphics [width=0.5\textwidth]{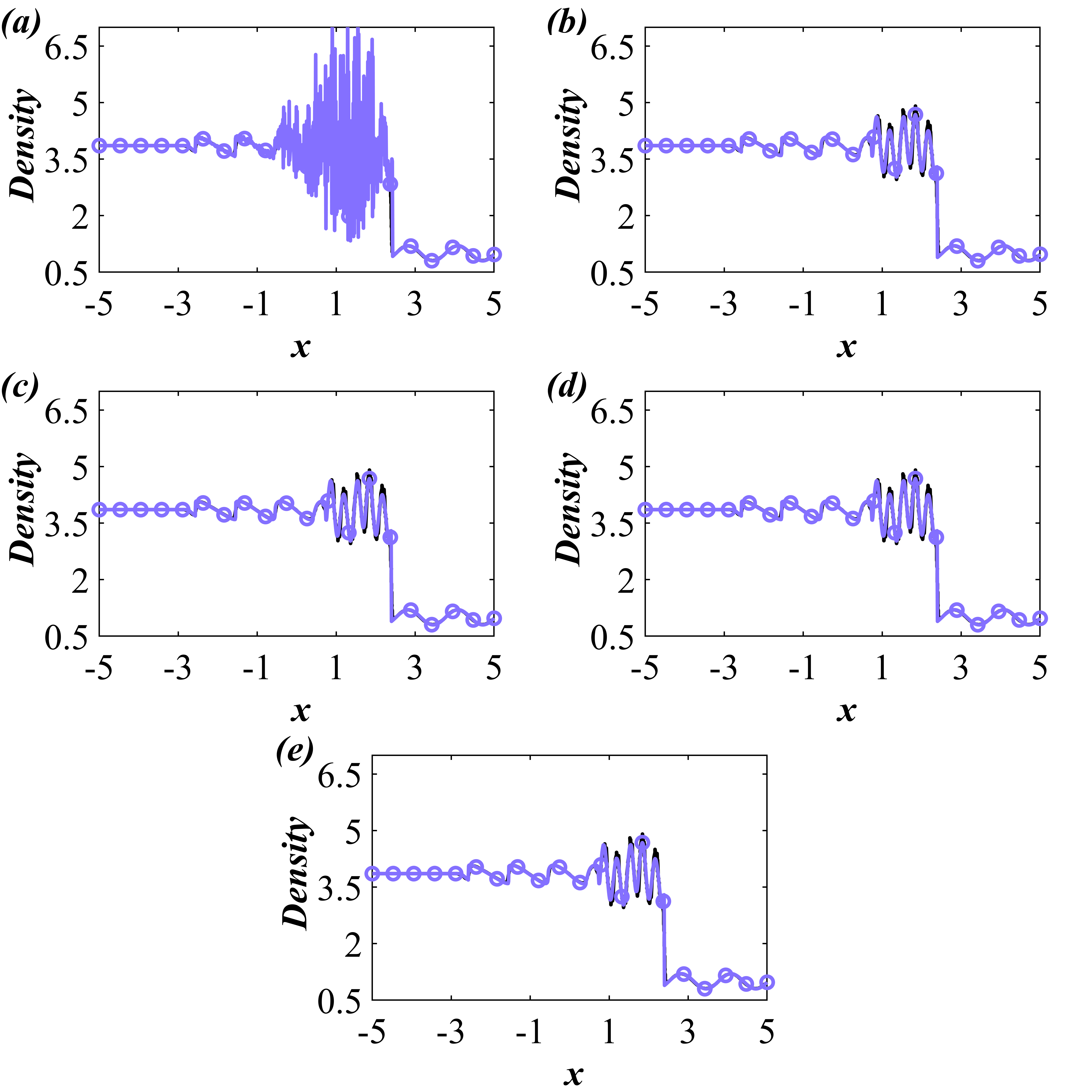}
\caption{Spatial density distribution at $t_f = 1.8$ for the Shu and Osher problem with different flux schemes. The solid black line represents the analytical solution \cite{shu1989efficient}, and the blue circles represent the numerical solution. (a) \textit{rhoCentralFoam} (b) HLLC-LM (d) HLLCP (d) AUSM+up (e) LDFSS}\label{Figure3}
\end{figure}

This problem represents an interaction of the Mach 3 shock wave with an entropy wave of sinusoidal nature, resulting in a large-scale disturbed flow field in the post-shock region. This test case was first proposed by Shu and Osher \cite{shu1989efficient} and was widely used by many research groups to test the accuracy of the higher-order interpolation schemes. This one-dimensional problem is analysed by considering the domain between [-5 5] and discretized with extremely fine resolution consisting of 8000 cells. This is intentionally done to analyse the solvers' behaviour in terms of their dispersive on the extremely refined grids. Both ends of the domain are applied with transmissive conditions. The domain is initialised with the following initial conditions. 

\[
    (\rho, u, p) =
    \begin{cases}
	(3.857143, 2.629369, 10.33333) & \text{if } x \leq -4, \\
	(1.0+ \xi sin (kx), 0, 1) & \text{if } x > -4.\\
    \end{cases}
\]
with $\xi = 0.2$ and $k=5$. The CFL number for this test is 0.2

Figure \ref{Figure3} demonstrates the density distribution at $t_f = 1.8$ over the length of the domain with the solutions overlapped with the reference numerical study \cite{shu1989efficient} that employs the same grid size as used in this investigation. The speculated spurious oscillations are quite severely amplified in this case for the solution of the default \textit{rhoCentralFoam} solver, even using a lower CFL number than the stability limit, while the other schemes produce decent, comparable results with the reference numerical study. Therefore, this test guarantees the robust behaviour of the time integration and implemented schemes used in this computation on the fine grid.

\subsubsection{Titarev-Toro problem}

\begin{figure}[t]
\centering
\includegraphics [width=0.5\textwidth]{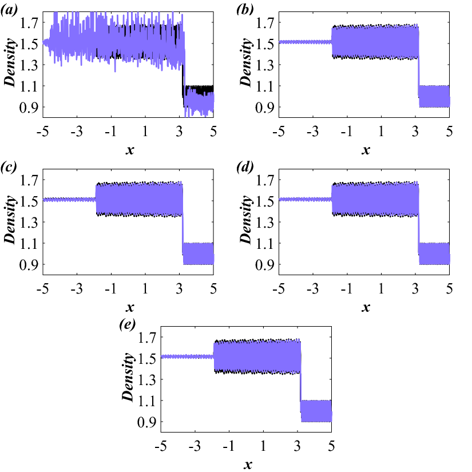}
\caption{Spatial density distribution at $t_f = 5.0$ for the Titarev-Toro problem with different flux schemes. The solid black line represents the solution \cite{titarev2004finite}, and the blue line represents the numerical solution. (a) \textit{rhoCentralFoam} (b) HLLC-LM (c) HLLCP (d) AUSM+up (e) LDFSS}\label{Figure4}
\end{figure}

This test is a severe version of the Shu and Osher problem \cite{li2023robust,titarev2004finite, titarev2005weno} where the shock wave interacts with an entropy wave with a higher frequency, resulting in extremely high fluctuating post-shock conditions. Once again, we employ the same computational domain with the same number of cells as used previously in the Shu and Osher problem; however, these initial conditions are used for this test.

\[
    (\rho, u, p) =
    \begin{cases}
	(1.515695, 0.5233346, 1.80500) & \text{if } x \leq -4.5, \\
	(1.0+ \xi sin (k\pi x), 0, 1) & \text{if } x > -4.5.\\
    \end{cases}
\]
with $\xi = 0.1$ and $k=20$. The solution is sought after $t_f = 5.0$

The default solver in \ref{Figure4} (a) for this case also shows a severe undulating, unphysical flow field in both post and pre-shock regions, which shows that the solver is extremely sensitive to any fluctuations, even if they are added \textit{ad-hoc}. Therefore, any numerical error arising from the calculation lag can be severely amplified by this solver, while the other schemes with the RK4 explicit time integration method show superior results, and they are in good faith with the reference numerical solution for all the schemes used in this study. Note that the solver preserves the density fluctuations provided in the initial conditions without any severe reduction/amplification in the amplitude ($\xi=0.1$) in the pre-shock region, while this is not applicable in the case of \textit{rhoCentralFoam} solver. 

\subsubsection{Two interacting strong blast waves}

\begin{figure}[t]
\includegraphics [width=0.5\textwidth]{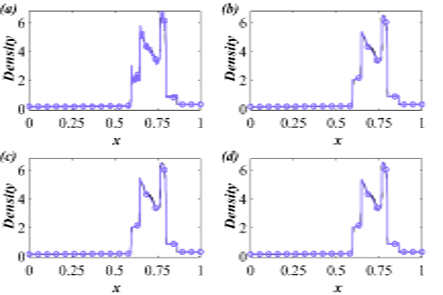}
\caption{Spatial density distribution at $t_f = 0.038$ for the blast wave interaction problem with different flux schemes. The solid black line represents the reference numerical solution \cite{woodward1984numerical,colella1984piecewise}, and the blue circles represent the numerical solution. (a) \textit{rhoCentralFoam} (b) HLLC-LM (c) HLLCP (d) AUSM+up}\label{Figure5}
\end{figure}

The dispersive nature of the default solver is also demonstrated in this test, where the \textit{ad-hoc} disturbances are not added in the initial conditions. This problem represents the complex evolution of the interaction of shock waves, which Woodward and Collela introduced \citep{woodward1984numerical,colella1984piecewise} for assessing the scheme's performance in resolving the interplay between various discontinuities of a strong nature. The domain spans [0 1] and is divided into 4000 cells of equal width. The higher resolution subjects the solver to extreme load in terms of its dispersive behaviour, especially for the default solver. The initial conditions for this case are as follows:

\[
    (\rho, u, p) =
    \begin{cases}
	(1, 0, 1000) & \text{if } x \leq 0.1, \\
	(1, 0, 0.01) & \text{if } 0.1 < x \leq 0.9, \\
        (1, 0, 100) & \text{if } x > 0.9.\\
    \end{cases}
\]
Although Woodward and Collela \cite{woodward1984numerical, colella1984piecewise} provide the solution's temporal evolution, the solution of the present numerical study is compared at the final time $t_f = 0.038$, with the time advancement achieved at CFL = 0.2.

Figure \ref{Figure5} compares the density solution for all the schemes used in this study. Once again, the spurious oscillations show up for the default solver \textit{rhoCentralFoam} even when the \textit{ad-hoc} disturbances are not added to the flow field as it is done for the previous two cases. This solver will evidently fail in resolving the smooth flow field on the extremely fine grid for the higher-dimensional problems. On the other hand, all the other implemented schemes show a fine match with the reference results in Ref. \cite{woodward1984numerical, colella1984piecewise}. Unfortunately, the LDFSS scheme diverged, producing negative temperature, showing trouble in resolving the high Mach number unsteady flow.

\begin{figure}[t]
\centering
\includegraphics [width=0.5\textwidth]{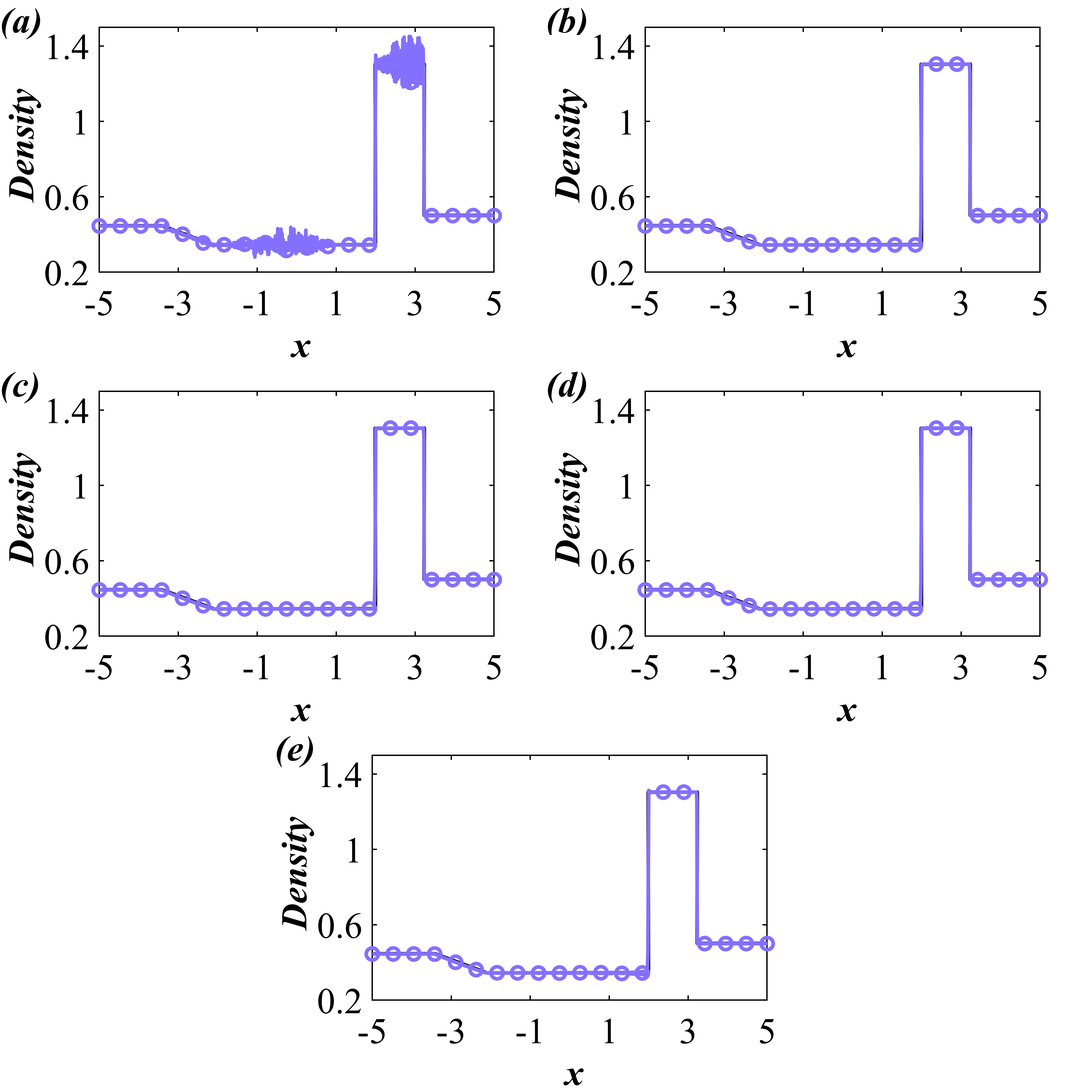}
\caption{Spatial density distribution at $t_f = 1.3$ for the Lax shock tube problem with different flux schemes. The solid black line represents the analytical solution \cite{lax2005weak}, and the blue circles represent the numerical solution. (a) \textit{rhoCentralFoam} (b) HLLC-LM (c) HLLCP (d) AUSM+up (e) LDFSS}\label{Figure6}
\end{figure}

\subsubsection{Lax shock tube}
The last problem in this series of one-dimensional tests is the Lax shock tube problem \cite{lax2005weak}, which we again solve using an extremely fine grid consisting of 4800 cells. The computational domain spans between [-5 5], and the initial conditions for this simulation are:
\[
    (\rho, u, p) =
    \begin{cases}
	(0.445, 0.698, 3.528) & \text{if } x \leq 0, \\
	(0.500, 0.000, 0.571) & \text{if } x > 0.\\
    \end{cases}
\]

Figure \ref{Figure6} displays the obtained solution using different schemes and the default solver's solution at $t=1.3$ using CFL = 0.5. Once again, for this case as well, spurious oscillations shoot up for the default solver, while the other schemes with RK4 integration show a smooth solution providing a decent match with the reference results. On the other hand, we corroborate the dispersive behavior of the LDFSS scheme by observing small-scale oscillations propagating upstream at $x=2$ (see inset).

\subsection{Two-dimensional tests}

\begin{figure}[h!]
\centering
\includegraphics [width=0.5\textwidth]{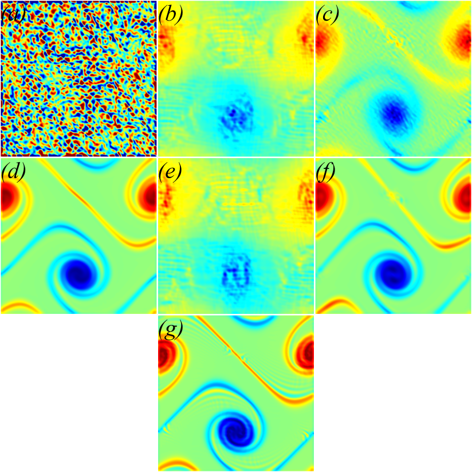}
\caption{Spatial density distribution at $t_f = 8$ for the double shear-layer problem with different flux schemes. (a) \textit{rhoCentralFoam} (b) HLLC2 (c) HLLC-LM (d) HLLCP (e) AUSM+up (f) LDFSS (g) \textit{rhoEnergyFoam}}\label{doubleShearLayer}
\end{figure}

So far, we have tested the schemes for the one-dimensional problems and conclude that \textit{rhoCentralFoam} solver produces unwanted spurious oscillations in the post-shock regions or regions with strong discontinuity on the extremely fine grid and using higher CFL numbers, while the other schemes produce decent results without admitting these numerical instabilities. However, a true picture can be revealed while simulating the two-dimensional problems where the information of the disturbance can now travel in two directions via the acoustic waves. The numerical scheme's tendency to cure the grid-aligned discontinuities fails due to inappropriate scaling of the acoustic and advective dissipation rates \cite{fleischmann2020shock} under the low Mach number limit. We carefully choose some of the standard benchmark test cases in two dimensions, which can generate strong grid-aligned discontinuities along with the cases where severe interface instabilites exist, all at different Mach numbers. The default solver and the implemented schemes are subjected to these tests, which may reveal these instabilities, at least for some of the schemes used in this study. This thorough test will enable a user to make an informed choice regarding the shock-capturing schemes that should be suitable for the problem under investigation.  

\subsubsection{Double shear-layer problem}
We begin this campaign of testing schemes for the two-dimensional flows under the low Mach number limit. To facilitate this, we consider a simple case of a double-shear layer problem, wherein two shear layers act as an interface between two opposite moving flows, and as a consequence, the shear layers \textit{roll-up} into strong vortices. It's simulation will shed light on the solvers/schemes' behaviour under the low Mach number conditions. We consider a square domain [x y] $\in$ [0 $2\pi$] $\times$ [0 $2\pi$] and discretize with 128 cells in each direction and apply with the periodic boundary conditions for the corresponding neighbour faces. These initial conditions are used to replicate the twin disturbed shear layers: 

\[
u(x, y) =
\begin{cases}
U_\infty \tanh\left( \dfrac{y - \pi / 2}{\delta_1} \right), & y \leq \pi \\
U_\infty \tanh\left( \dfrac{3\pi / 2 - y}{\delta_1} \right), & y > \pi
\end{cases}
\]
\[
v(x, y) = \delta_2 \sin(x)
\]
\[
\rho = 1.0, \quad
p = \dfrac{1}{\gamma M_\infty^2}
\]
\[
U_\infty = 1.0, \quad
\delta_1 = \dfrac{\pi}{15}, \quad
\delta_2 = 0.05
\]

We consider the Mach number of the flow to be 0.01, representing the incompressible regime of the flow and obtain the solution at $t_f = 8$, where we expect a large roll-up of the shear layer \cite{ishiko2006implicit,kitamura2016reduced,chen2020hllc+}. The solution proceeds with a Courant number (of CFL number) of 0.2 for all the schemes, except for the case of the LDFSS scheme, which is two orders of magnitude lower to maintain the stability of this scheme. We also show the solution of the high-fidelity solver \textit{rhoEnergyFoam} operated in Mode A state \cite{modesti2017low} as the reference solution. Figure \ref{doubleShearLayer} is the outcome of this numerical test. The default solver fails to retain vortices, but the uncorrected HLLC scheme (Figure \ref{doubleShearLayer} (b)) and the AUSM+up scheme Figure \ref{doubleShearLayer} (e) at least show the vortex cores, but are highly smeared. On the other hand, the HLLC-LM (Figure \ref{doubleShearLayer} (c)), HLLCP (Figure \ref{doubleShearLayer} (d)), and LDFSS Figure \ref{doubleShearLayer} (f) schemes resolve them; however, the HLLC-LM scheme shows diffused structures along with numerical noises at certain locations of the flow field. The HLLCP and LDFSS schemes show comparable results that match to a great extent with the reference solution obtained using the \textit{rhoEnergyFoam} solver. This test essentially points to the HLLCP and LDFSS scheme's correct scaling of the acoustic signals at very low Mach numbers, but a significant cost is incurred in terms of the number of iterations required to reach the desired end time for the LDFSS scheme.

\subsubsection{Two-dimensional Riemann problems}
We continue the discussion for the Riemann problem, but in two dimensions. For this, we consider a computational domain [x y] that ranges across [0 1] $\times$ [0 1] with the Initial Boundary Value Problem (IBVP) across four interfaces. Many configurations are available based on different solutions proposed by Lax \cite{lax1998solution}. 

\begin{figure}[h!]
\centering
\includegraphics [width=0.5\textwidth]{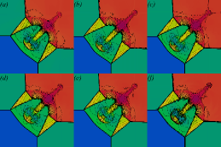}
\caption{Spatial density distribution at $t_f = 0.3$ for the two-dimensional Riemann problem (configuration 3) with different flux schemes. (a) \textit{rhoCentralFoam} (b) HLLC-LM (c) HLLCP (d) AUSM+up (e) LDFSS (f) \textit{rhoEnergyFoam}}\label{config3}
\end{figure}

These test cases produce only one elementary wave (shock wave, expansion wave, or contact) at the interface of the Riemann problem, unlike the previous one-dimensional test cases, where all three waves originated from the initial discontinuity location. This section is dedicated to numerically simulating configuration 3, which produces a jet-like structure while the shock wave propagates and intersects at a triple point, and the slip line through this point entrains inside the jet. The slip lines for this configuration experience Kelvin-Helmholtz Instability (KHI); therefore, it can be an ideal problem for assessing the dissipative nature of the schemes considered in this investigation. The reader is redirected to Ref. \cite{chamarthi2023wave} for the reference high-resolution numerical solution. Further, we also consider configuration 6, which is the representation of four interfaces with slip lines. The conclusions derived from this test case can strengthen the understanding of the scheme's characteristics. We specifically also compare the simulated cases with another high-fidelity solver developed by Modesti and Pirozolli \cite{modesti2017low}. 

The computational domain is divided into 1024 cells in each direction, and the solution is evolved with a CFL of 0.2. Initial conditions $(\rho, u, v, p)$ for all the regions are given below for configurations 3 and 6. Transmissive conditions are used at all boundaries of the domain to avoid reflections.

Configuration 3:
\[
	(\rho, u, v, p) =
	\begin{cases}
		(1.5, 0, 0, 1.5) & \text{if } x > 0.5, y > 0.5, \\
		(0.5323, 1.206, 0, 0.3) & \text{if } x \leq 0.5, y > 0.5, \\
		(0.138, 1.206, 1.206, 0.029) & \text{if } x \leq 0.5, y \leq 0.5, \\
		(0.5323, 0, 1.206, 0.3) & \text{if } x > 0.5, y \leq 0.5. \\
	\end{cases}
\]

Configuration 6:
\[
	(\rho, u, v, p) =
	\begin{cases}
		(1, 0.75, -0.5, 1) & \text{if } x > 0.5, y > 0.5, \\
		(2, 0.75, 0.5, 1) & \text{if } x \leq 0.5, y > 0.5, \\
		(1, -0.75, 0.5, 1) & \text{if } x \leq 0.5, y \leq 0.5, \\
		(3, -0.75, -0.5, 1) & \text{if } x > 0.5, y \leq 0.5. \\
	\end{cases}
\]

\begin{figure}[h!]
\centering
\includegraphics [width=0.5\textwidth]{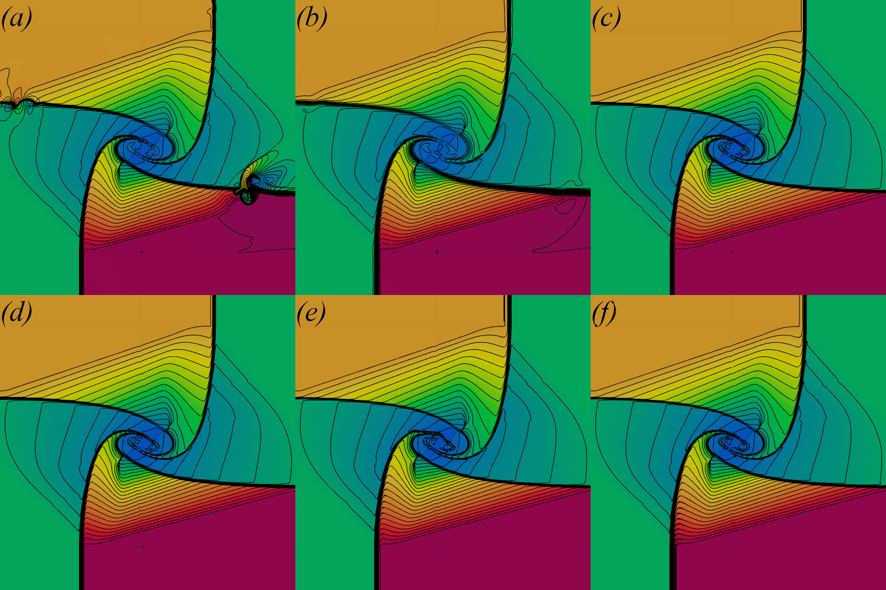}
\caption{Spatial density distribution at $t_f = 0.3$ for the two-dimensional Riemann problem (configuration 6) with different flux schemes. (a) \textit{rhoCentralFoam} with van Leer reconstruction (b) \textit{rhoCentralFoam} with minmod reconstruction (c) HLLC-LM (d) HLLCP (e) AUSM+up (f) LDFSS}\label{config6}
\end{figure}


The 51 evenly spaced isopycnics between 0.5 and 1.7 at $t_f = 0.3$ are displayed in Figure \ref{config3} for the configuration 3 solution. The findings corresponding to the reference solution, the high-fidelity solver \textit{rhoEnergyFoam}, are displayed in Figure \ref{config3} (f). This grid size resolves complex features. The solution found using the default solver is shown in Figure \ref{config3} (a). It can generate the general flow features rather well, but the contact (slip line) is diffuse and has no discernible KHI across it. However, the HLLC family (see Figure \ref{config3} (b) and (c)) schemes with more closely spaced iso-lines improve the resolution of the slip lines. For these schemes, the reference solution provides strong evidence for the generation of small-scale fluid dynamic instabilities above the slip line. Conversely, the results of the LDFSS scheme (Figure \ref{config3} (e)) demonstrate well-developed large-scale amplitudes of KHI exhibiting the dissipation characteristics comparable to the reference solution, whilst the AUSM+up scheme (Figure \ref{config3} (d)) likewise maintains the small-scale instability over the slip line.



We examine configuration 6 suggested by Lax \cite{lax1998solution}, for the following two-dimensional Riemann problem, in which every interface rotates clockwise around the domain's center. Figure \ref{config6} (a) displays the results of our operation of the standard solver using the KNP shock-capturing strategy with van Leer reconstruction. The artifact of artificial instability that appears in the solver's solution process is an artificially disturbed slip line. By employing this mix of techniques, it is hypothesized that the solver may diverge, resulting in significantly increased numerical noise. The solution is stabilized, though, using the same shock capturing scheme but a slightly more diffuse reconstruction scheme, minmod (see Figure \ref{config6} (b)). This is achieved at the cost of increased numerical diffusion, which is reflected in the resolution of the slip line and smeared slip line filament at the domain's center. With lengthy, elongated slip lines at the mixing location, the other schemes yield results that are quite equivalent to the standard solution given in Ref. \cite{lax1998solution}, demonstrating less artificial diffusion in the schemes that are put into practice.


In conclusion, these two-dimensional Riemann problems basically demonstrate that, in terms of the resolution of the slip line, the OpenFOAM framework's default solver generates extremely diffuse solutions. The diffusive reconstruction strategy can reduce the numerical oscillatory behavior that is visible in configuration 6, but the other approaches yield results that are on par with those obtained with the high-fidelity solver \textit{rhoEnergyFoam}.

\subsubsection{Strong shock diffraction around a corner}

\begin{figure}[h!]
\centering
\includegraphics [width=0.5\textwidth]{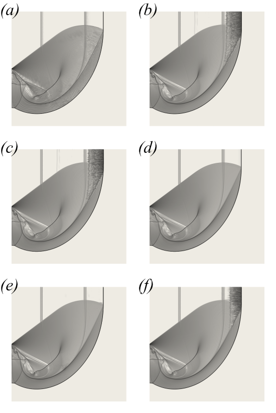}
\caption{Numerical schlieren images using density gradient magnitude contour at $t_f = 0.157$ for the shock diffracting around a sharp corner with different flux schemes represented on the log scale. (a) \textit{rhoCentralFoam} (b) HLLC1 (c) HLLC2 (d) HLLC-LM (e) HLLCP (f) AUSM+up}\label{Figure8}
\end{figure}

This is the classical test case designed by Quirk \cite{quirk1997contribution} to test the low-dissipation Riemann solvers, wherein the author showed the failure of the schemes to produce stable results. Additionally, the flow solution reveals complex flow patterns that can be ideally used to assess the dissipation characteristics of the adopted numerical methodology. 

To simulate this problem, we adopt the setup suggested in Ref. \cite{fleischmann2020shock}. The domain spans in x and y direction between [0 1] $\times$ [0 1] with 1280 cells in each direction. We consider a strong shock wave with a Mach number of 5.09 to be diffracting around the corner. The computational domain is initialised with $(\rho, u, v ,p)$ =(1 0 0 1/1.4) throughout. Transmissive boundary conditions are used for all the boundaries except the patch at $x = 0$ and $0.5 \leq y \leq 1$ is prescribed with the post-shock conditions corresponding to Mach 5.09. The solution at $t_f = 0.157$ (or 0.8/Ma) is used for comparing the results corresponding to different schemes/solvers. We also compare the results of the classical HLLC scheme using two different formulations (see Section \ref{Section2}) (without implementing any correction) to facilitate a direct comparison with the solutions obtained using the respective corrected versions (HLLC-LM and HLLCP), showing the effectiveness of these implemented corrections to cure the grid-aligned shock instabilities.

Figure \ref{Figure8} portrays the numerical schlieren image at $t_f =0.157$. Although the default solver in Figure \ref{Figure8} (a) provides a stable solution, we observe high undulation in the flow field, especially behind the strong diffracting shock wave, but on a positive note, the grid-aligned shock instabilites are absent. Moreover, the slip line originating around the diffracting corner is diffuse with apparently no KHI. On the other hand, the HLLC1 (Figure \ref{Figure8} (b)) and HLLC2 (Figure \ref{Figure8} (c)) schemes also produce strong numerical noise specifically behind the grid-aligned portion of the diffracting shock wave. These observations are consistent with the findings reported by Fleischmann et al. \cite{fleischmann2020shock} for the same computational setup. This eventually supports that the implementation of the HLLC schemes in the original form and the central form is indeed correct. We also observe slip line \textit{roll-ups} describing the KHI which is otherwise not resolved by the default solver. On the contrary, these instabilities vanish when using the corrected version of the HLLC schemes, i.e., HLLC-LM (Figure \ref{Figure8} (d)) and HLLCP (Figure \ref{Figure8} (e)). The results remain consistent for the resolution of the slip line using the AUSM+up (Figure \ref{Figure8} (f)) scheme as well, but severe disturbances in the flow still remain, pointing to its inefficiency for resolving the grid-aligned discontinuity correctly; however, the spatial extent is smaller compared to the basic HLLC scheme. Unfortunately, the LDFSS scheme fails to provide stable results, producing floating-point error during a few initial iterations.

\subsubsection{Sedov blast wave}

We consider another classical problem of the Sedov blast wave as suggesting in Ref. \cite{fleischmann2020shock, hu2013positivity, tasker2008test}. This is an extremely difficult problem to solve, which can pose a serious challenge to the numerical flux schemes, as it involves the blast wave's propagation with an extremely high pressure ratio initially localised in a small region of space expanding in the presence of a near vacuum. It is a classical problem that finds application in the supernova explosion in space; therefore, a standard problem in astrophysics \cite{hu2024shock, tasker2008test}. Note that some portion of this blast wave perfectly aligns with the grid, therefore, this case serves as an ideal case for judging the ability of the schemes to avoid grid-aligned instabilities behind the strong moving blast wave. The domain for this problem [x y] ranges across [-1.2 1.2] $\times$ [-1.2 1.2] and is discretised with 960 cells in each direction. We apply the following initial conditions with non-reflecting boundary conditions for all the boundaries.

\[
(\rho, u, v, p) =
\begin{cases}
(1,\ 0,\ 0,\ 3.5 \times 10^5), & \text{if } \sqrt{x^2 + y^2} < 0.005, \\
(1,\ 0,\ 0,\ 10^{-10}), & \text{otherwise}
\end{cases}
\]
The simulation lasts till $t_f = 0.1$ by imposing the Courant number of 0.2. 

\begin{figure}[h!]
\centering
\includegraphics [width=0.5\textwidth]{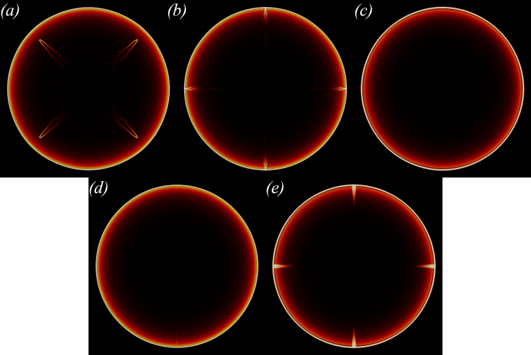}
\caption{Spatial density distribution at $t_f = 0.3$ for the Sedov blast wave problem with different flux schemes. (a) \textit{rhoCentralFoam} (b) HLLC1 (c) HLLC-LM (d) HLLCP (e) AUSM+up}\label{sedov}
\end{figure}

The \textit{rhoCentralFoam} in Figure \ref{sedov} (a) shows the density gradient magnitude, highlighting small-scale, unphysical instabilities which aren't though grid-aligned. However, the standard HLLC scheme shows these instabilities in Figure \ref{sedov} (b), while they vanish in its corrected versions (HLLC-LM in Figure \ref{sedov} (c) and HLLCP in Figure \ref{sedov} (d)). The AUSM+up Figure \ref{sedov} (e) scheme also shows these instabilities, while the LDFSS scheme crashes, highlighting its inability to resolve strong discontinuities. Overall, the solution possesses the same order of accuracy for all the schemes concerning the blast wave's propagation front. 

\subsubsection{Flow over cylinder at Mach 3 and 20}

\begin{figure}[h!]
\centering
\includegraphics [width=0.5\textwidth]{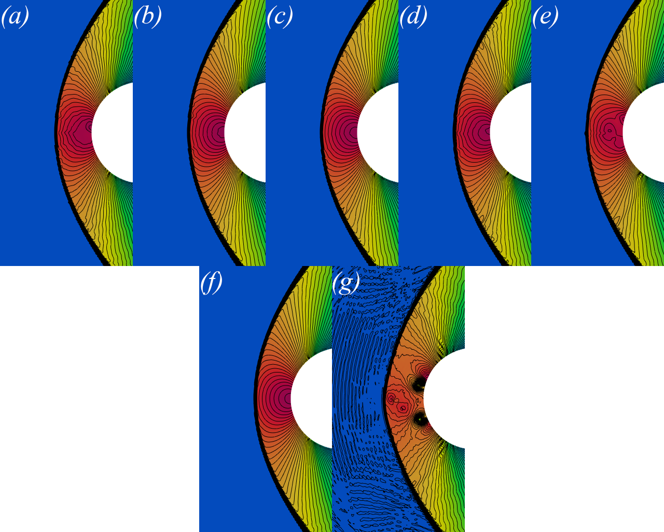}
\caption{Spatial density distribution at $t_f = 1.5$ for the supersonic flow over the cylinder problem at Mach 3 with different flux schemes. (a) HLLC2 (b) HLLC-LM (c) HLLCP (d) AUSM+up (e) LDFSS (f) \textit{rhoCentralFoam} (g) \textit{rhoEnergyFoam}}\label{cylinderMa3}
\end{figure}

We now consider Mach 3 and 20 supersonic flow over the cylinder, which is known to produce \textit{carbuncle shock} phenomena generating a severely distorted bow shock at the stagnation line of the cylinder \cite{quirk1997contribution}. We consider these two Mach numbers for this simulation, which serves as a stringent test for the schemes to generate a physically meaningful bow shock profile and stagnation region for both Mach numbers. The higher Mach number helps determine the scheme's and solver's stability and robustness. For the Mach 3 case, we consider a cylindrical inlet of diameter 2D, where D = 0.2 is the cylindrical diameter with its centre coinciding with the center of the cylinder. The rest of the ends are connected with the outlet, which is assumed to be non-reflective owing to supersonic flow. Mathematically, the domain can be described as $-2D\leq r\leq-0.5D$ (with the centre for the cylinder fixed at the origin) and $-\pi/2\leq\theta\leq\pi/2$. This domain is discretised with 80 cells in the azimuthal direction and 160 cells in the radial direction, generating 25,600 cells, with a maximum aspect ratio of around 4. Here, the radial coordinate, r, is measured from the centre of the cylinder, while the azimuthal coordinate's ($\theta$) origin is at the stagnation point. We use the same non-dimensionalization method to represent the fluid properties in the non-dimensional form. The inlet is prescribed with $(\rho, u, v, p)$ =(1.4 Ma 0 1) where Ma = 3 or 20 depending on the case considered. The Courant number is set to 0.2 for both Mach numbers

Figure \ref{cylinderMa3} portrays the solution obtained from the various schemes used in this investigation for the Mach 3 case using the 51 isobars between 1 and 12 at $t_f = 1.5$. We note that none of the schemes suffer from carbuncle instability and generate physically consistent results except the solution of the reference solution of the high-fidelity solver \textit{rhoEnergyFoam} (Figure \ref{cylinderMa3} (g)). However, key differences exist among the solutions that were obtained. Figure \ref{cylinderMa3} (a) shows the results of the classical HLLC correction showing severely asymmetric flow at the stagnation point, while their corrected versions do not show these anomalies. The corrections balance the local acoustic propagation speeds and the convection speed under the low Mach number limits (behind the bow shock), therefore, provide correct scales of diffusion. The AUSM+up scheme in \ref{cylinderMa3} (d) produces comparable results with the HLLC scheme but with slight disturbances behind the curved portion of the bow shock wave, while the LDFSS scheme in \ref{cylinderMa3} (e) suffers from severe instabilities at the stagnation point. The default solver (Figure \ref{cylinderMa3} (f)) provides physically consistent results, but the shock wave is slightly \textit{smeared}.   

\begin{figure}[h!]
\centering
\includegraphics [width=0.5\textwidth]{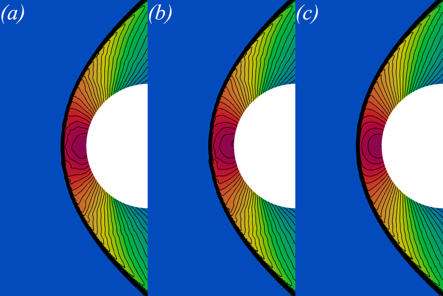}
\caption{Spatial density distribution at $t_f = 0.5$ for the hypersonic flow over the cylinder problem at Mach 20 with different flux schemes. (a) \textit{rhoCentralFoam} (b) HLLC2 (c) HLLC-LM}\label{cylinderMa20}
\end{figure}

Likewise, Figure \ref{cylinderMa20} shows 31 isobars for between 1 and 512, showcasing their behaviour at hypersonic Mach numbers at $t_f = 0.5$. Only HLLC (Figure \ref{cylinderMa20} (a)), HLLC-LM (Figure \ref{cylinderMa20} (b)) schemes, and the default solver (Figure \ref{cylinderMa20} (c)) generate results at this high Mach number flow, while the other schemes fail during the initial iterations with a floating-point exception. None of the schemes suffers from the carbuncle problem, but the HLLC scheme shows a tiny recirculating flow feature at the stagnation point of the cylinder. The HLLC-LM scheme shows slightly better results, showcasing its effectiveness in resolving the correct dissipative behaviour at the low Mach number limit, while the HLLCP scheme fails, as stated earlier. The other schemes based on the flux splitting method are unstable and diverge at this high Mach number flow. In a nutshell, this exercise shows that the HLLC-LM scheme is robust in resolving the stagnation flow of the cylinder while maintaining the correct shock waves even at a high Mach number of 20 with minimum numerical diffusion.


\subsubsection{Kelvin-Helmholtz instability}

\begin{figure}[h!]
\centering
\includegraphics [width=0.5\textwidth]{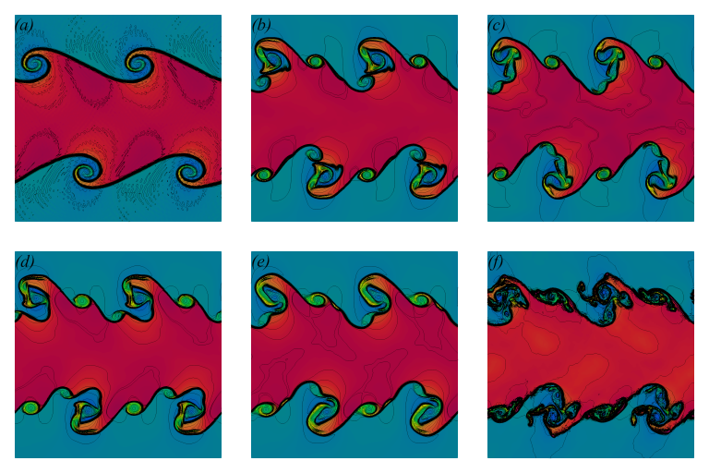}
\caption{Spatial density distribution at $t_f = 0.8$ for the Kelvin-Helmholtz instability problem with different flux schemes. (a) \textit{rhoCentralFoam} (b) HLLC-LM (c) HLLCP (d) AUSM+up (e) LDFSS (f) \textit{rhoEnergyFoam}}\label{KHI}
\end{figure}

This problem can be the ideal candidate to test the fidelity and stability of the scheme, where we artificially impart perturbations to the flow through initial conditions \cite{chamarthi2023wave}. Although we tested the scheme's ability to resolve the KHI through various test cases as previously highlighted, where the KHI naturally evolved over the slip line, this test case is an ideal problem to see the behaviour of the schemes in the presence of the perturbed flow field resolved on a fine grid. We consider the domain spanning between [x y] that ranges across [0 1] $\times$ [0 1], divided into 1024 cells in each direction, and we apply these initial conditions \cite{chamarthi2023wave}:

\[
\rho(x, y) =
\begin{cases}
2, & \text{if } 0.25 < y \leq 0.75, \\
1, & \text{otherwise}
\end{cases}
\]

\[
u(x, y) =
\begin{cases}
0.5, & \text{if } 0.25 < y \leq 0.75, \\
-0.5, & \text{otherwise}
\end{cases}
\]

\[
\begin{split}
v(x, y) &= 0.1 \sin(4\pi x) \left[
\exp\left( -\frac{(y - 0.75)^2}{2\sigma^2} \right) \right] \\ 
&\qquad + \left[\exp\left( -\frac{(y - 0.25)^2}{2\sigma^2} \right) \right] 
\end{split}
\]

\[
\sigma = \frac{0.05}{\sqrt{2}}
\]

\[
p(x, y) = 2.5
\]
The left and right boundaries are periodic with non-reflecting boundary conditions for the remaining boundaries. We seek the solution at $t_f = 0.8$.


Figure \ref{KHI} shows the density contour with 31 superimposed isopycnics between 0.1 and 0.8. along with the high-fidelity solution of the \textit{rhoEnergyFoam} solver for the reference numerical solution. The KHI is manifested by all the solvers; however, we observe subtle differences among them. The spurious oscillations show up for the default solver while they are absent for the other schemes used in this investigation. The roll-up of the slip line is perfectly symmetric in the default solver, while the other schemes show secondary roll-ups matching the fidelity of the reference solution. The closest match is obtained in the case of the HLLCP scheme, showcasing its excellent dissipation characteristics. The next comparable scheme is the HLLC-LM, showcasing results similar to those of the HLLCP scheme. The inherent wave modelling embodied in the HLLC family of schemes, while calculating the numerical fluxes, explains these observations. The AUSM+up  and LDFSS schemes are slightly dissipative compared to the HLLC-type schemes; however, it's very minuscule.

\subsubsection{Richtmyer-Meshkov instability}


The next problem in this series is the interaction of the shock wave with a perturbed interface of the fluids with different densities. This setup finds an application in enhancing mixing between two fluids \cite{chamarthi2023wave, chamarthi2023gradient}. The interaction of the shock wave results in a hydrodynamic instability termed Richtmyer-Meshkov instability (RMI). To simulate this problem, we consider a computational domain [x y] between [0 4] $\times$ [0 1] divided into 2560 cells in the x direction and 640 cells in the y direction. The left and right boundaries are applied with Dirichlet conditions, while the top and bottom boundaries are imposed with periodic boundary conditions. These initial conditions are imposed inside the domain:

\[
(\rho, u, v, p) =
\begin{cases}
(5.04,\ 0,\ 0,\ 1), \\ \qquad \qquad \quad \text{if } x < 2.9 - 0.1 \sin\left[ 2\pi (y + 0.25) \right], \\ \\
(1,\ 0,\ 0,\ 1), \\ \qquad \qquad \quad \text{if } x < 3.2, \\ \\
\left(1.4112,\ -\frac{665}{1556},\ 0,\ 1.628\right), \quad \\ \qquad \qquad \quad \text{otherwise}
\end{cases}
\]

\begin{figure}[h!]
\centering
\includegraphics [width=0.5\textwidth]{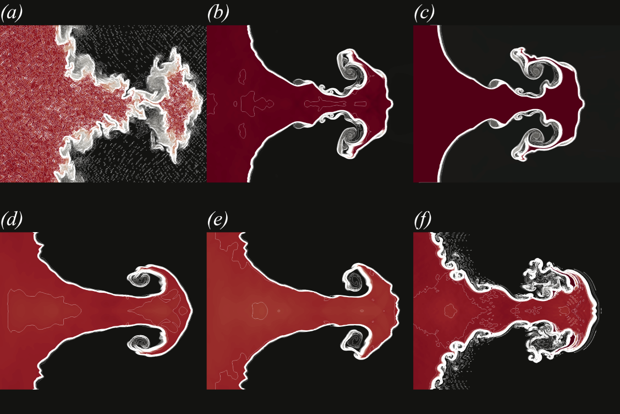}
\caption{Spatial density distribution at $t_f = 0.8$ for the Richtmyer-Meshkov instability problem with different flux schemes. (a) \textit{rhoCentralFoam} (b) HLLC-LM (c) HLLCP (d) AUSM+up (e) LDFSS (f) \textit{rhoEnergyFoam}}\label{RMI}
\end{figure}

The solution of the several numerical techniques used in this investigation is shown in Figure \ref{RMI} at $t_f = 9$. Again, it should come as no surprise that the default solver generates significant false numerical noise, but it does capture the flow properties to a certain extent. However, AUSM+up (Figure \ref{RMI} (d)) and LDFSS (Figure \ref{RMI} (e)) are slightly dissipative with regard to the resolution of the interface, with AUSM+up resolving the interface better than the LDFSS scheme. In contrast, the HLLC family (HLLC-LM Figure \ref{RMI} (b) and HLLCP Figure \ref{RMI} (c)) schemes exhibit rich structures with small-scale instabilities over the neck and head of the resolved structure. As seen in Figure \ref{RMI} (f), the reference higher-order numerical solution produced by the rhoEnergyFoam solver comes closest to the fidelity of the HLLC-type schemes.



\subsubsection{Axisymmetric case: Muzzle blast}

The solvers or schemes for axisymmetric issues, including intricate Mach reflection characteristics with a vortex, supersonic core flow, and microscopic vortexlets, can be put to the test using this problem. This problem had been previously reported by Wang and Widhopf \cite{wang1990numerical} and Batten et al. \cite{batten1997choice} using numerical computation, and its corresponding experiments are available in Ref. \cite{schmidt1985noise}.  With a diameter of D = 0.15, a lip thickness of D/2, and a length of D, the shock wave with a Mach number of 1.76 diffracts around the lip. The range of the computational domain [x y] is [0 6D] × [0 3D] for simulating this test case. The origin of the nozzle's axis, which is on the x-axis, corresponds to the global coordinate system. A structured grid is used to discretize the domain so that $\delta x = 0.005$ and $\delta y = 0.005$. To replicate this numerically, the post-shock conditions are applied so that the normal shock wave is exactly near the muzzle's exit.


\begin{figure}[h!]
\centering
\includegraphics [width=0.5\textwidth]{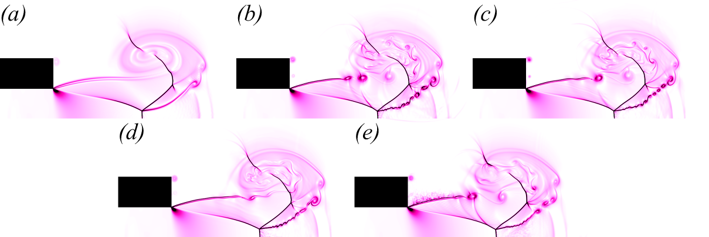}
\caption{Density gradient magnitude at $t_f = 1.5 $ ms for the muzzle blast problem with different flux schemes. (a) \textit{rhoCentralFoam} (b) HLLC-LM (c) HLLCP (d) AUSM+up (e) LDFSS}\label{muzzle}
\end{figure}


The solution at $t_f = 1.5 $ ms is featured in Figure \ref{muzzle}. The Mach stem at the axis, the primary supersonic vortex, the shock waves inside it, and various triple points are distinctly visible in the simulated results. The default solver produces mean flow features with relatively smooth slip lines, whereas the other schemes show rich flow structures involving small-scale vortexlets, especially over the slip line originating from the triple point of the Mach stem. KHI over the slip line originating from the nozzle lip is pronounced in the HLLC family schemes with large-scale amplitude over it, and perhaps its breakdown into tiny vortexlets while it is engulfed into the primary vortex core. The AUSM+up scheme shows comparable fidelity with the HLLC-type schemes, while the LDFSS scheme shows severe artificial perturbation near the slip line at the same location. It is worthwhile to note that the LDFSS scheme is unstable with van Leer reconstruction; therefore, we use a slightly dissipative reconstruction scheme, minmod, which explains the slightly lower fidelity compared to the other implemented schemes.

\subsubsection{Supersonic flow over circular bump}

\begin{figure}[h!]
\centering
\includegraphics [width=0.5\textwidth]{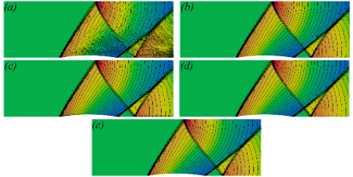}
\caption{Spatial density distribution at steady state for the supersonic flow over a circular bump. (a) rhoCentralFoam (b) HLLC-LM (c) HLLCP (d) AUSM+up (e) LDFSS}\label{bump_contour}
\end{figure}

The final test in this series of two-dimensional flow assessment is the supersonic flow over the circular bump with a 4 \% circular arc. This test case poses a stringent challenge for the numerical schemes to produce stable results behind a multiple-shock system along with the supersonic flow expanding over the circular arc. The computational domain and the associated dimensions can be referred to, from Ref. \cite{oliani2023icsfoam}, with a fine grid topology consisting of 10800 cells. However, unlike their study, in which they used realistic values of the flow properties, we rely on the hypothetical gas whose product of specific heat ratio and characteristic gas constant is unity to maintain consistency throughout this article. The inlet is prescribed with the Dirichlet boundary conditions with $(\rho, U, p) = (1.4, 1.4, 1)$ corresponding to a free stream Mach number of 1.4. Since the flow is supersonic throughout, we simply extrapolate the value from the internal cells to the cells at the outlet boundary. The top and bottom patches are applied with reflecting boundary conditions. The domain is initialised with properties of the same values as those used to describe the inlet properties. We use the Courant number 0.5 to instigate the inherent instabilities in the numerics; however, we generate stable results for all the schemes used in this study, except the default solver at this Courant number. It is worthwhile to note that the default solver produces stable results at least up to Courant number 0.25, as demonstrated in the work of Oliani et al \cite{oliani2023icsfoam}.  


Figure \ref{bump_contour} demonstrates the results for this numerical test. It is now no surprise that the default solver in the OpenFoam framework cannot generate results without introducing spurious oscillations in the solution, and therefore, there are severely amplified perturbations in the flow field for the results corresponding to this solver. On the contrary, all the other schemes produce smooth stable flow fields irrespective of the schemes and they compare qualitatively well with the numerical results of SU2 \cite{economon2016su2}, a well-validated open-source solver, as shown in Figure 9 of Ref. \cite{oliani2023icsfoam}. We observe a small-amplitude perturbed flow field originating over the bottom surface in the case of the HLLCP scheme (Figure \ref{bump_contour} (c)) over the circular bump surface and its aft region, indicating numerically dispersive behaviour similar to the findings reported in Figure 33 of Ref. \cite{kim1998improvement}.


\begin{figure}[h!]
\centering
\includegraphics [width=0.5\textwidth]{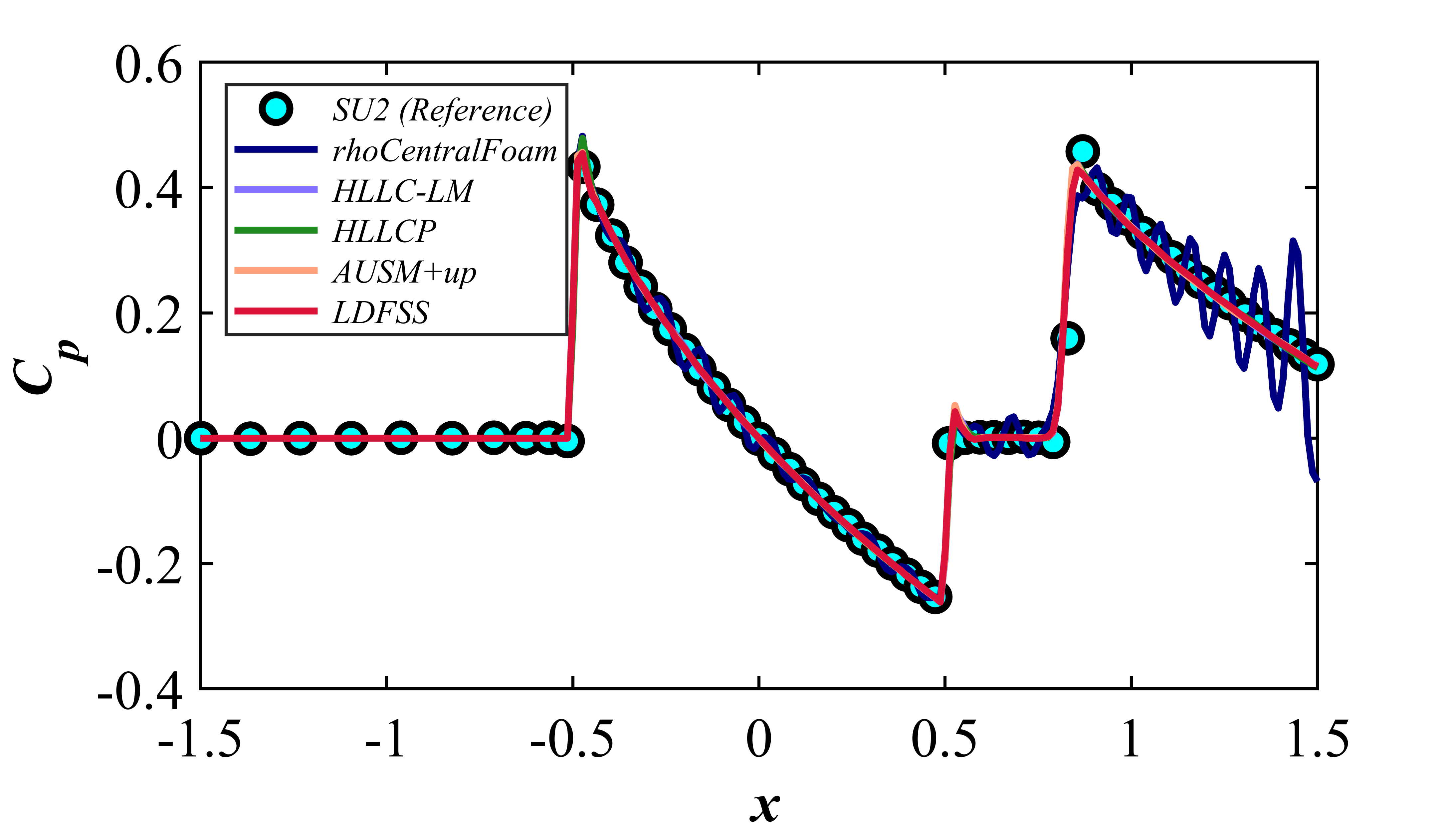}
\caption{Variation of pressure coefficient, $C_p$ at steady state for different solvers and schemes on the lower wall of the supersonic bump flow.}\label{bump_plot}
\end{figure}


We also compare the data produced with the same grid size but utilizing the SU2 framework, which is provided in Ref. \cite{oliani2023icsfoam}, with the pressure distribution in non-dimensional form in the current study. These findings are displayed in Figure \ref{bump_plot}, where the default solver's evident perturbations are evident. However, the other methods yield respectable and comparable results to the reference numerical solution that was produced using the SU2 suite. Although there is a significant pressure spike at $x = 0.5$  behind the second shock wave, the stable values surpass the peak of the SU2 pressure distribution at the first shock located at the bump's root. The pressure distribution at the third shock, which originates as the reflection of the incident shock from the top wall, is marginally underestimated by all the schemes. In comparison to the other techniques, the default solver significantly underpredicts, demonstrating its inefficiency in modeling the shock wave with a highly oscillatory solution after this point. However, no post-shock oscillations are seen in their solution since they simulated this case with a low Courant number (0.25). Even the solution obtained using \textit{rhoCentralFoam} in the reference work of Oliani et al. \cite{oliani2023icsfoam}, obtained using minmod limiter, shows severe underprediction consistent with the current results of this solver.


These thorough two-dimensional analyses demonstrate that the default solver might not be appropriate for high-fidelity calculations (for instance, when using an extremely fine grid). If it is, the user must pay a significant penalty by using drastically reduced Courant numbers to suppress these spurious instabilities. In the following section, we provide closure to the evaluation of the Euler equations by illustrating the behavior of various solvers and schemes for three-dimensional problems.

\subsection{Three-dimensional tests}

\subsubsection{Taylor-Green vortex}

\begin{figure}[h!]
\centering
\includegraphics [width=0.5\textwidth]{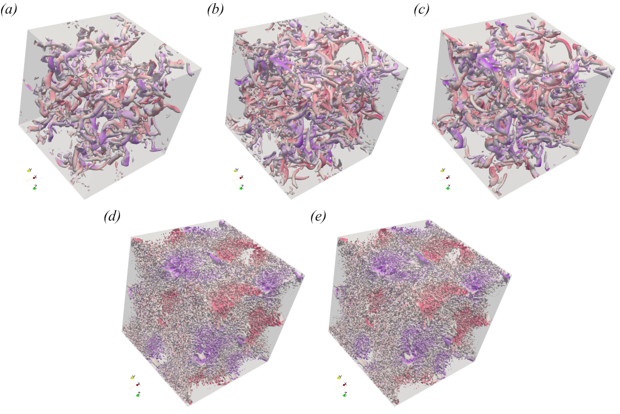}
\caption{Iso-surface of Q criterion (Q = 2) at $t_f = 10$ for the Taylor-Green vortex problem with different flux schemes. (a) HLLC-LM (b) HLLCP (c) LDFSS (d) \textit{rhoEnergyFoam}, mode A (e) \textit{rhoEnergyFoam}, mode B}\label{TGV}
\end{figure}

\begin{figure}[h!]
\centering
\includegraphics [width=0.5\textwidth]{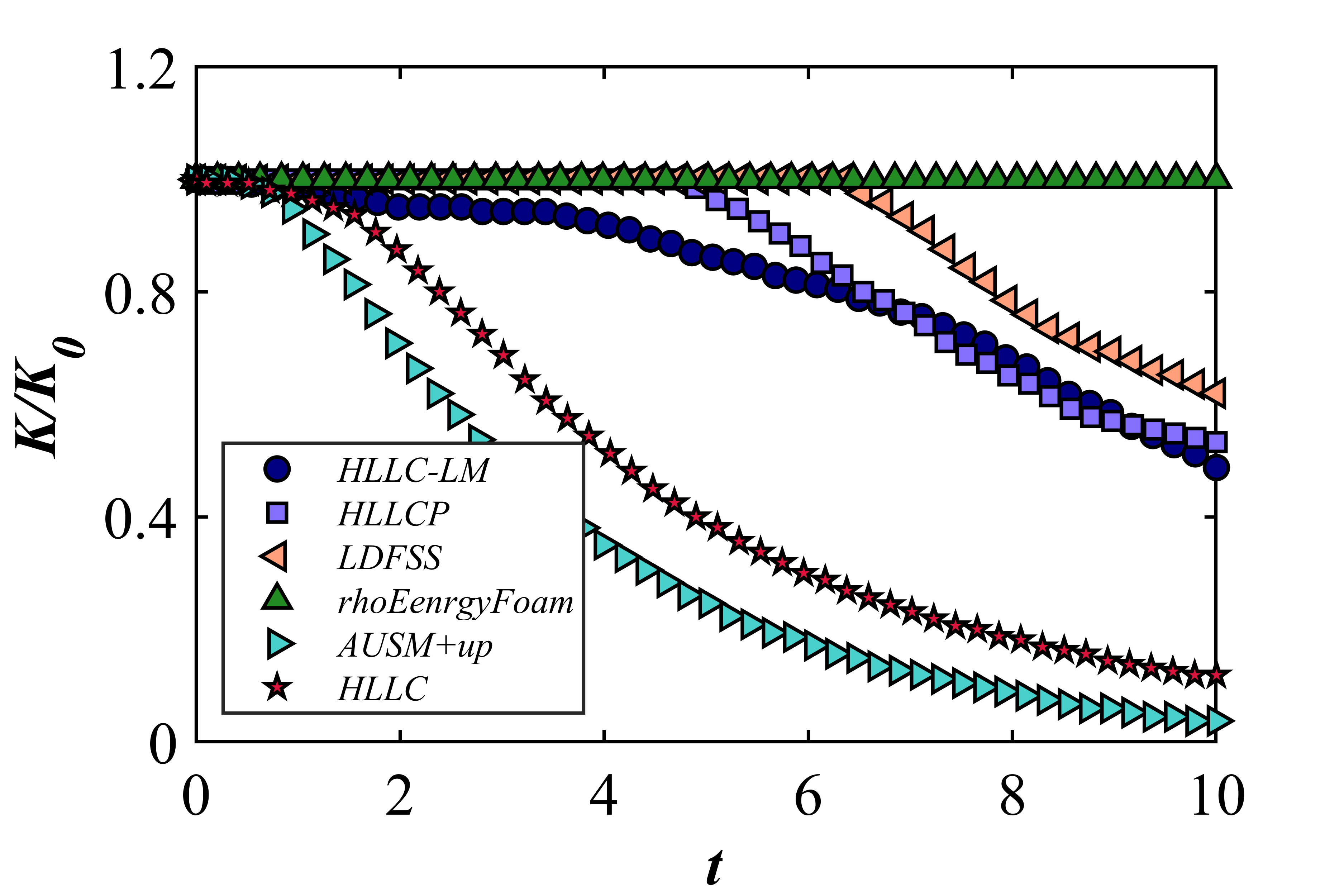}
\caption{Temporal evolution of the turbulent kinetic energy with time for the Taylor-Green vortex with different flux schemes. }\label{TGV_plot}
\end{figure}

For assessing the dissipation characteristics of the three-dimensional inviscid flow, we consider a popular test case of the Taylor-Green vortex. This is a standard case that shows the decay of the large-scale vortex over time by the action of the artificial numerical viscosity inherent in the schemes. In general, the ideal scheme should preserve the total kinetic energy, but it can decay owing to the numerical viscosity. The better schemes show slow decay of the kinetic energy while retaining the small-scale structures \cite{chamarthi2023wave}.


We consider a cubical, triple periodic, computational domain with each side of length $2\pi$, discretised with $128^3$ cells. We impose these initial conditions inside the domain corresponding to Mach 0.08 \cite{chamarthi2023wave} :

\[
\begin{pmatrix}
\rho \\
u \\
v \\
w \\
p
\end{pmatrix}
=
\begin{pmatrix}
1 \\
\sin x \cos y \cos z \\
- \cos x \sin y \cos z \\
0 \\
100 + \dfrac{ \left[ \cos(2z) + 2 \right] \left[ \cos(2x) + \cos(2y) \right] - 2 }{16}
\end{pmatrix}
\]

The reference high-fidelity solution of the \textit{rhoEnergyFoam} solver operating in modes A and B for the reference solution is shown in Figure \ref{TGV}, along with the results of the schemes that maintain prominent structures represented by the iso-surface of Q-criterion of 2 colored with the x component of the velocity after the final time $t_f = 10$. The other schemes entirely dissipated the large-scale vortical structures, but only the HLLC-LM, HLLCP, and LDFSS displayed them. The fidelity displayed by the rhoEnergyFoam solver is not matched by any of the methods. The HLLCP and LDFSS schemes are the least dissipative, according to the qualitative assessment of the number of structures retained, with the HLLC-LM scheme offering somewhat greater diffusion levels.


\begin{figure}[h!]
\centering
\includegraphics [width=0.5\textwidth]{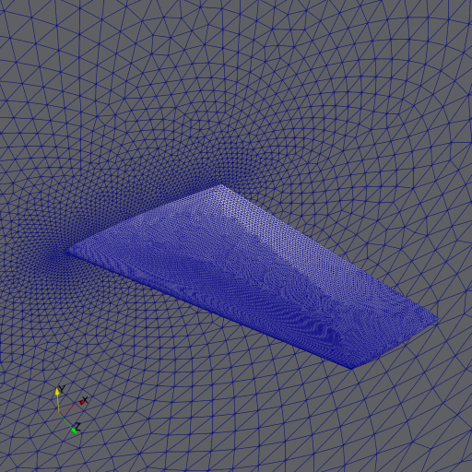}
\caption{Closeup view of a typical unstructured surface mesh over the ONERA-M6 wing }\label{onera_mesh}
\end{figure}

\begin{figure}[h!]
\centering
\includegraphics [width=0.5\textwidth]{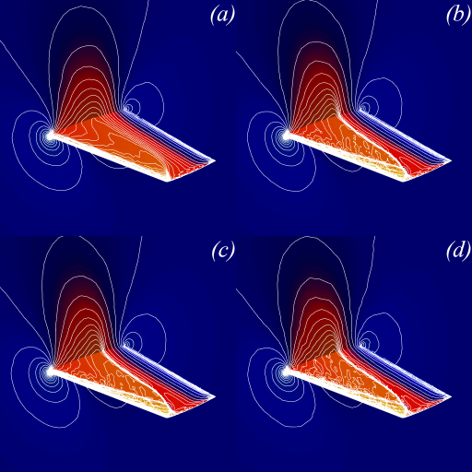}
\caption{Spatial distribution of the pressure contour at steady state for different flux schemes. (a) \textit{rhoCentralFoam} (b) HLLC-LM (c) AUSM+up (d) LDFSS }\label{onera_pressure}
\end{figure}


Figure \ref{TGV_plot} shows the normalised total kinetic energy evolution with time. The \textit{rhoEnergyFoam} perfectly retains all the kinetic energy with negligible dissipation, thus serving as the benchmark for the other schemes under investigation. The standard HLLC scheme is excessively dissipative as mentioned earlier; however, its corrected version performs better under the low Mach number limit, thus retaining at least the large-scale structures as shown in Figure \ref{TGV} (a) and Figure \ref{TGV} (b). The Kinetic energy decay is the most severe for the AUSM+up scheme, pointing out its non-suitability for the flow under a very low Mach number limit. The energy level of the LDFSS scheme is slightly higher than the rest of these two schemes, which essentially points out its non-dissipative nature in the weakly compressible flow limit. It is important to note that the LDFSS scheme operates stably under very low CFL (Courant number) conditions, which is at least 20 times smaller than the CFL number used for the other schemes. 
\subsubsection{Transonic flow over ONERA-M6 wing}

\begin{figure}[h!]
\centering
\includegraphics [width=0.5\textwidth]{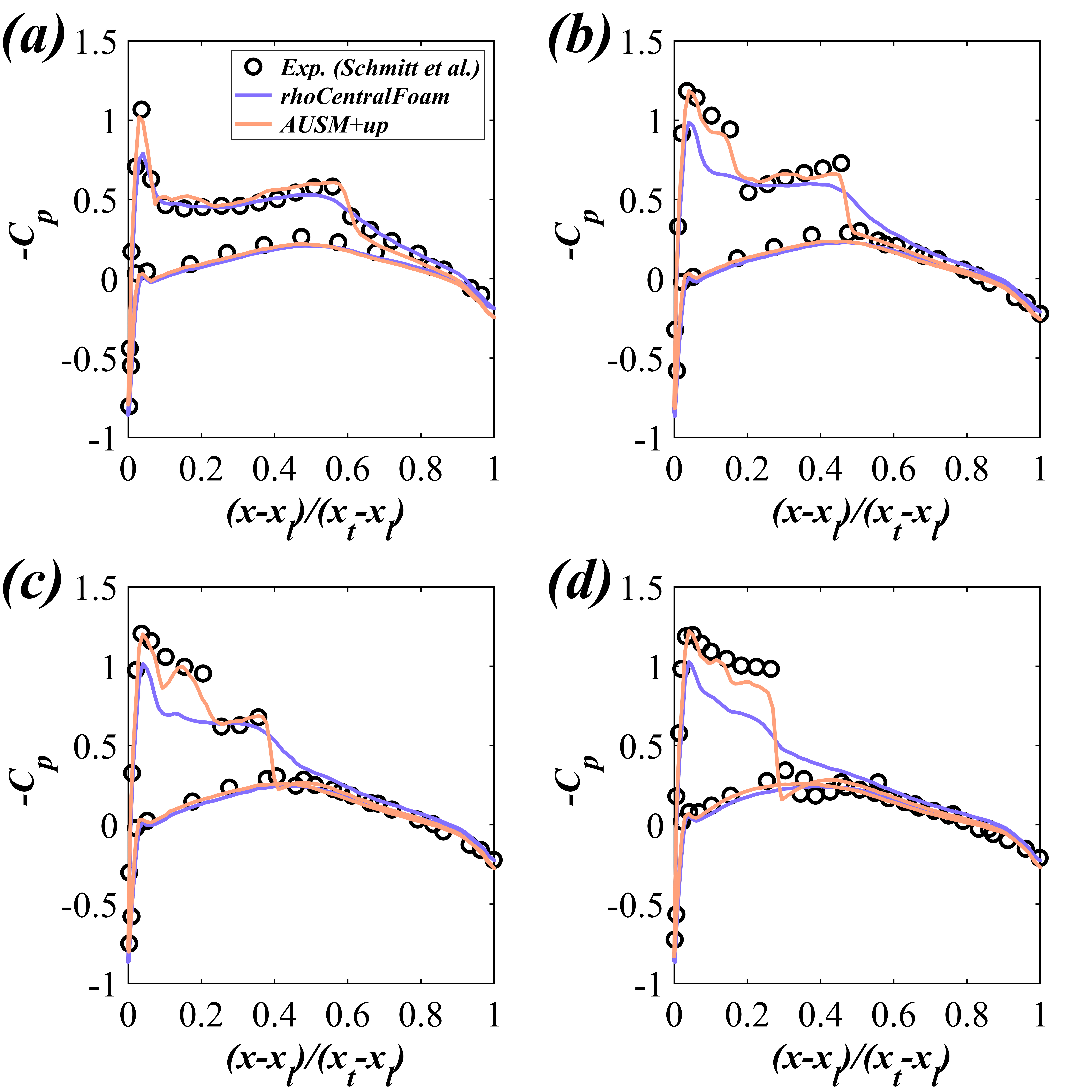}
\caption{Coefficient of pressure distribution, $C_p$ comparison obtained with the default solver and AUSM+up scheme over different wing spans. (a) $z/b=0.2$ (b) $z/b=0.65$ (c) $z/b=0.8$ (d) $z/b=0.9$ }\label{onera_plot}
\end{figure}

We now consider the last problem in this series, which assesses the solver or schemes' capability to generate results on the unstructured grid, showcasing the solver's implementation versatility in the OpenFOAM framework. We consider the ONERA-M6 wing at a free stream Mach number of 0.8395 at an angle of attack, $\alpha = 3.06^\circ$. We replicate the setup provided by Modesti and Pirozolli \cite{modesti2017low}, which used the domain size [x y z] = [0 10c] $\times$ [0 10c] $\times$ [0 5c], with c as the chord of the section located at the wing's root, discritzed with 341797 tetrahedral cells as shown in Figure \ref{onera_mesh}. A Courant number of 0.5 is used for all the schemes under investigation except for the case of the default solver, where we restrict it to 0.2 to ensure its stable operation.


Figure \ref{onera_pressure} shows the pressure contour superimposed with 31 isolines between $0.4p_{\infty}$ and $1.585p_{\infty}$ over the wing's surface for some of the schemes under investigation. It clearly shows that the implemented schemes outperform the standard OpenFOAM solver. The lambda shock structure is sharply resolved for all the schemes at all the wing span locations, except it's excessively smeared concerning the \textit{rhoCentralFoam} solver. We also perform a quantitative comparison with the experimental data concerning the surface pressure coefficient provided by Schmitt and Charpin \cite{schmitt1979pressure} at different wing sections. We only show the AUSM+up scheme and the \textit{rhoCentralFoam} for a clear representation of the comparison. The other implemented schemes produce almost the same results with minute variations among them. Figure \ref{onera_plot} (a) - (d) shows these distributions at $z/b=0.2, 0.65, 0.8$ and 0.9, respectively, where $b$ is the span of the wing. The pressure rise over the top surface representing the primary shock foot of the lambda shock at $z/b =0.2$ is barely visible, while the AUSM+up scheme shows a sharp rise achieving the experimental Cp values; moreover, the pressure distribution perfectly aligns with the experimental values for this scheme. The agreement between the default solver and the experimental values progressively worsens on moving downstream along the wing span, and it is severely underpredicted at $z/b = 0.9$ where the feet of the lambda shock wave merge.

\section{Results: Solution of Navier-Stokes equations\label{Section4}}

\subsection{Deep subsonic, laminar flow over cylinder at Re = 200}

\begin{figure}[h!]
\centering
\includegraphics [width=0.5\textwidth]{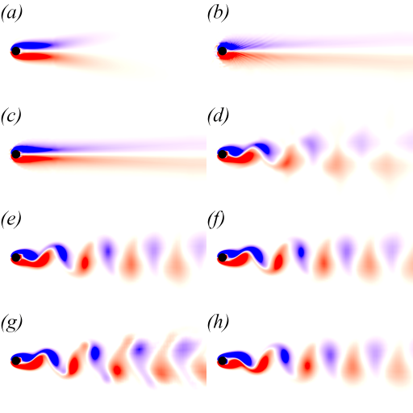}
\caption{Von-Karman vortex street behind the cylinder at Reynolds number 200 obtained at $t = 5 $ ms for the different flux schemes. (a) \textit{rhoCentralFoam} (b) AUSM+up (c) HLLC (d) HLLC-LM (e) HLLCP (f) LDFSS (g) rhoEnergyFoam (h) \textit{pimpleCentralFoam}}\label{vortexShedding}
\end{figure}

We choose a standard fluid dynamic problem of two-dimensional flow over a cylinder at a Reynolds number of 200 to begin the examination of the solver's performance with physical viscosity consideration. The work of Kraposhin et al. \cite{kraposhin2018hybrid} served as the inspiration for this challenge. The authors have demonstrated the hybrid solver's capacity to resolve the incompressible flow in the low Mach number limit in this reference. Their hybrid technique uses a typical incompressible solver called \textit{pimpleFoam} in the OpenFOAM framework, which devices the PIMPLE algorithm in conjunction with the algorithm of the standard compressible solver taken into consideration in this study. To address the low Mach number flow, the researchers employed a sensor that modifies the algorithm according to the local Mach number. They considered solving the laminar flow over a cylinder at Re = 200 to demonstrate the capabilities of their built solver. The alternate vortex shedding from the cylinder's top and bottom sides is the subject of this classical problem, which is a perfect example of how the scheme behaves diffusively in the deep-subsonic zone of the flow regime. We use the exact same computational setup that was used by Kraposhin et al. \cite{kraposhin2018hybrid} using the free stream velocity ($U_{\infty}$) of 10 m/s with the dynamic viscosity of $18.5 \mu Pas$. The diameter of the cylinder is 0.157 mm, which sets the Reynolds number $(Re = \rho_{\infty} U_{\infty} D/\mu_{\infty})$ of the flow to 200. We further impose standard atmospheric conditions for free-stream pressure and temperature by prescribing them as $p_{\infty} = 101325 Pa$ and $T_{\infty} = 300 K$ at the inlet. The free stream Mach number ($M_{\infty} = U_{\infty}/a_{\infty}$), based on this free-stream properties is 0.029. The simulation lasts till t = 5 ms, which covers approximately 50 cycles of lift fluctuation (or vortex shedding cycles) over the cylinder. The viscous terms in the Navier-Stokes equations are discretised with the central scheme of second-order accuracy \cite{greenshields2010implementation}.

\begin{table*}[t]
\centering
\caption{Comparison of aerodynamic coefficients and Strouhal number for different solvers and schemes used in this investigation, along with the reference studies. \label{table1}}
\resizebox{\textwidth}{!}{%
\begin{tabular}{lclcccc}
\toprule
\textbf{ } & \textbf{Elements} & \textbf{Methodology} & $\bar{C}_d$ & $C_d'$ & $C_l'$ & \textit{St} \\
\midrule
HLLC-LM & 19400 & Implemented scheme & 1.398 & 1.172 & 3.013 & 0.152 \\
HLLCP & 19400 & Implemented scheme & 1.374 & 0.00751 & 0.245 & 0.162 \\
LDFSS & 19400 & Implemented scheme & 1.370 & 0.00740 & 0.248 & 0.162 \\
rhoEnergyFoam & 19400 & FVM , density-based solver & 1.372 & 0.0447 & 0.25349 & 0.167 \\
pimpleCentralFoam & 19400 & Hybrid solver & 1.352 & 0.0066 & 0.258 & 0.160 \\
Liang et al. \cite{liang2009high} & 1336 & Spectral differences & 1.365 & 0.0086 & 0.232 & 0.164 \\
Sharman et al. \cite{sharman2005numerical} & 14441 & FVM with collocated storage, SIMPLE & 1.330 & 0.0064 & 0.230 & 0.164 \\
Meneghini et al. \cite{meneghini2001numerical} & 13696 & FEM & 1.370 & -- & -- & 0.165 \\
Kang \cite{kang2003characteristics} & 62127 & Immersed boundary & 1.330 & -- & 0.320 & 0.165 \\
Roshko experiment \cite{roshko1954development} & -- & -- & -- & -- & -- & 0.16 -- 0.17 \\
Zahm report \cite{zahm1927flow} & -- & -- & 1.25 -- 1.4 & -- & -- & -- \\
\bottomrule
\end{tabular}%
}
\end{table*}

Figure \ref{vortexShedding} shows the wake structures (using spanwise vorticity component) resolved by the schemes employed in this study, along with the solution concerning the \textit{rhoCentralFoam} solver. We also show the reference solution obtained using the hybrid solver and the high fidelity solver \textit{rhoEnergyFoam} at t = 5 ms when the flow is completely developed. As explained earlier, it is expected that the vortices shed from the alternate sides of the cylinder at this Reynolds number; however, the default solver in Figure \ref{vortexShedding} (a) shows the mean wake structure with a stable separation bubble behind the cylinder. This observation sheds light on the dissipative behaviour of the scheme and the inability of the solver to resolve the flow in low Mach number flow. The characteristics are shown by the AUSM+up scheme and the uncorrected version of the HLLC scheme as well, thus pointing out its dissipative characteristics under the low Mach number limit. However, we see a conspicuous wake structure representing the well-known Von Karman vortex street for the corrected versions of the HLLC-family schemes and the LDFSS scheme. Although the wake structure is resolved for the HLLC-LM scheme, the shed vortices are smeared, while the HLLCP scheme shows distinct vortices with clear resolution. Comparable resolution is also obtained using the LDFSS scheme. These analyses essentially point to the relatively dissipative nature of the HLLC-LM corrected version under the low Mach number limit, while the solutions of the HLLCP and LDFSS schemes are in line with the reference solution. It is important to note that the LDFSS scheme is stable using the Courant number approximately 20 times smaller than the Courant number used for the rest of the schemes (CFL = 0.2), consistent with the simulation of the Taylor-Green vortex using the Euler equations. On the other hand, using the higher CFL number leads to the diverged solution.

\begin{figure}[h!]
\centering
\includegraphics [width=0.5\textwidth]{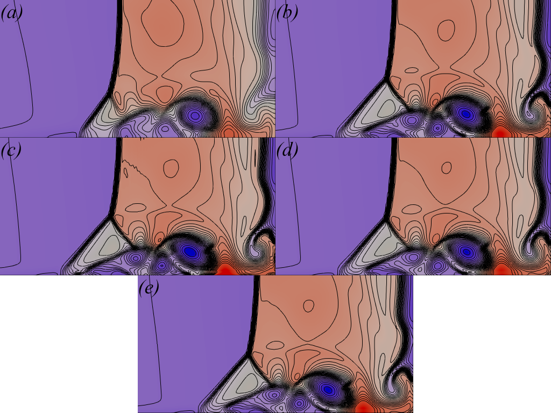}
\caption{Spatial density distribution at $t = 1$ for viscous shock tube problem at Reynolds number 200 using a coarse grid obtained using different flux schemes. (a) \textit{rhoCentralFoam} (b) HLLC-LM (c) HLLCP (d) AUSM+up (e) LDFSS}\label{vs_re_200_coarse}
\end{figure}

Table \ref{table1} summarises the key observations of the unsteady features of the vortex shedding phenomena along with some of the key reference studies with different methodologies employed. We only report the observations for the schemes that provide dynamically correct behaviour while results from the other schemes are not shown. This table compares the drag coefficient $\bar{C_d}$ and the Root Mean Square (RMS) of the drag and lift coefficients $C_d'$ and $C_l'$ respectively along with the Strouhal number $(St = fD/U_{\infty})$, providing an overall idea of the signals captured in an unsteady signal. Although, wake structure of the cylinder in the case of the HLLC-LM scheme is resolved, $\bar{C_D}$ is significantly higher than all the schemes and the reference studies as well. Ideally, the root mean square of the drag coefficient should be a small number, but it is significantly higher owing to significant numerical noise associated with the scheme, which also amplifies the mean coefficient. The $C_l'$ is also significantly higher while the Strouhal number is severely underpredicted than the reference simulation in the present investigation and the values reported in the reference studies, suggesting its strong dispersive behaviour under an extremely low Mach number limit. The other schemes, such as HLLCP and LDFSS schemes provide reasonable agreement with the reference values for all the quantities involved.


\subsection{Viscous Riemann problem: Re = 200 and 2500}

\begin{figure}[h!]
\centering
\includegraphics [width=0.5\textwidth]{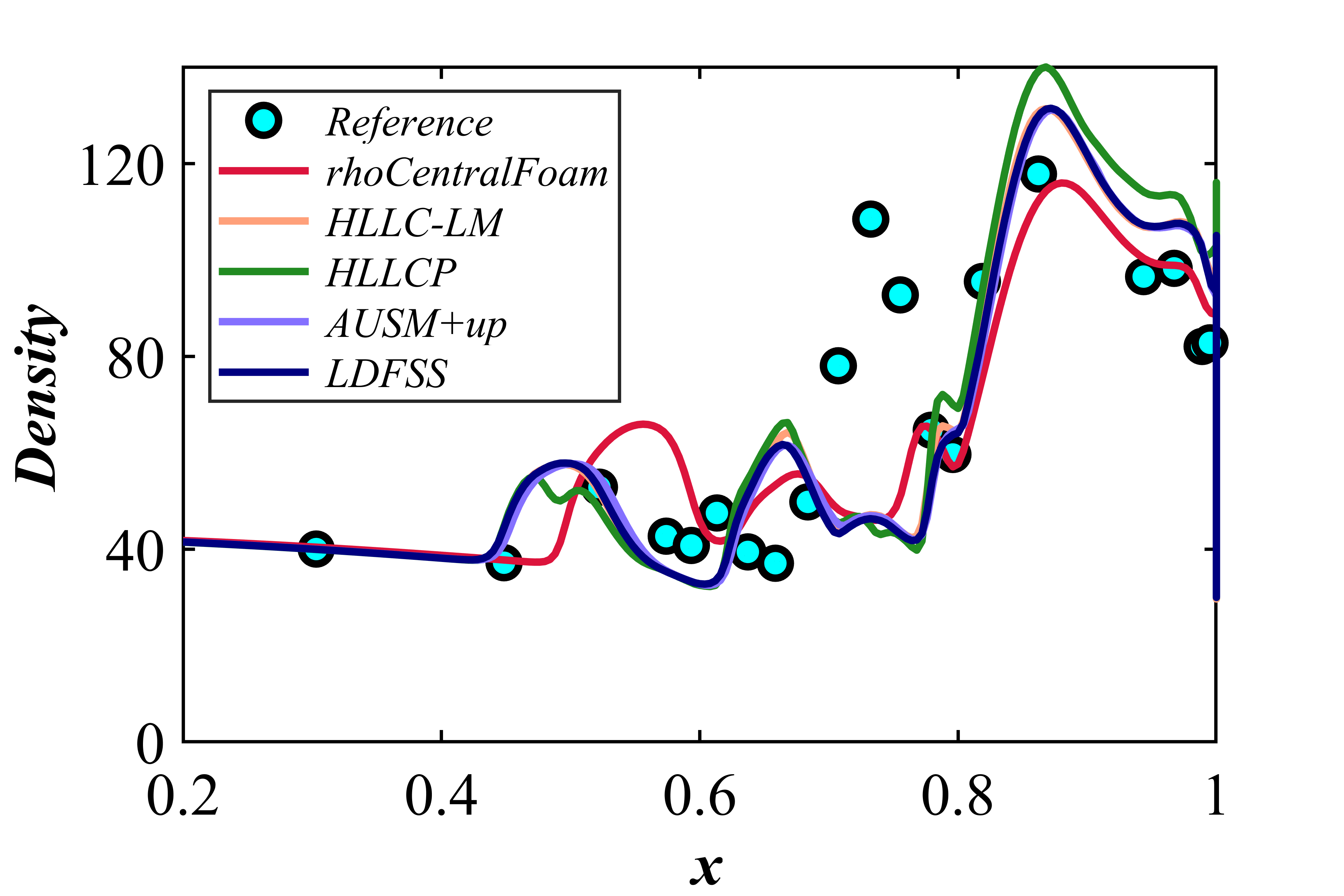}
\caption{Comparison of the density distribution over the bottom wall of the shock tube at $t = 1$ for the viscous shock tube problem at Reynolds number 200 using a coarse grid obtained using different flux schemes. }\label{vs_re_200_coarse_plot}
\end{figure}

The second problem in this series is the viscous shock tube problem. This problem will help assess the characteristics of the shock-capturing schemes in the presence of physical viscosity, which is a more realistic scenario for fluid flow. A right propagating shock front is initialised inside a shock tube of the domain [x y] $\in$ [0 1] $\times$ [0 0.5], as suggested by Daru et al. \cite{daru2009numerical}. The shock wave propagates towards the right end of the tube, inducing flow behind the incident shock wave, reflects off the right wall, and interacts with the boundary layer, which develops while the shock wave propagates towards the right. While interacting with the boundary layer, the shock wave separates it, forming a lambda shock foot, and vortices are shed off from the shear layer. It is noted that the resolution of the primary vortex shed depends on the shock-capturing scheme used and the grid resolution. The primary vortex significantly elongates in the transverse direction, provided the grid resolution is converged and the scheme employed is accurate and non-dissipative \cite{fukushima2024improved}.

\begin{figure}[h!]
\centering
\includegraphics [width=0.5\textwidth]{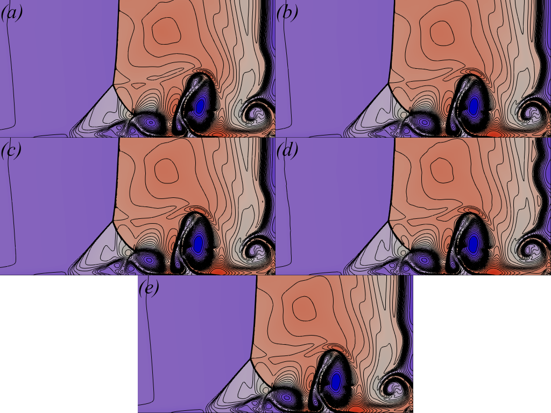}
\caption{Spatial density distribution at $t = 1$ for viscous shock tube problem at Reynolds number 200 using a fine grid obtained using different flux schemes. (a) \textit{rhoCentralFoam} (b) HLLC-LM (c) HLLCP (d) AUSM+up (e) LDFSS}\label{vs_re_200_fine}
\end{figure}

We solve the Navier-Stokes equation for this simulation on a grid size that is extremely coarse $(\delta x = \delta y = 0.004)$ and on a fine mesh with 8 times the resolution of this coarse mesh. The coarse mesh can help to analyse the true dissipative behaviour, while the fine mesh can justify the dispersive behaviour of the schemes. The viscous terms are discretised using the central scheme of second-order accuracy. Reynolds number (Re) is set at 200 (through viscosity), and the Prandtl number (Pr) is 0.7. The top patch is imposed with a symmetry boundary condition, as only the bottom half of the shock tube is considered in this investigation. All the other remaining boundary is applied with no-slip, adiabatic wall boundary conditions. $\gamma$ is 1.4 for this set of simulations. Initial conditions $(\rho, u, v, p)$ are imposed based on the following conditions:
	
	\[
	(\rho, u, v, p) =
	\begin{cases}
		(120, 0, 0, 120/\gamma) & \text{if } x \leq 0.5, \\
		(1.2, 0, 0, 1.2/\gamma) & \text{if } x > 0.5.\\
	\end{cases}
	\]
CFL for this case is imposed at 0.2 for this case as well. The end time for this solution is $t_f = 1$.

\begin{figure}[h!]
\centering
\includegraphics [width=0.5\textwidth]{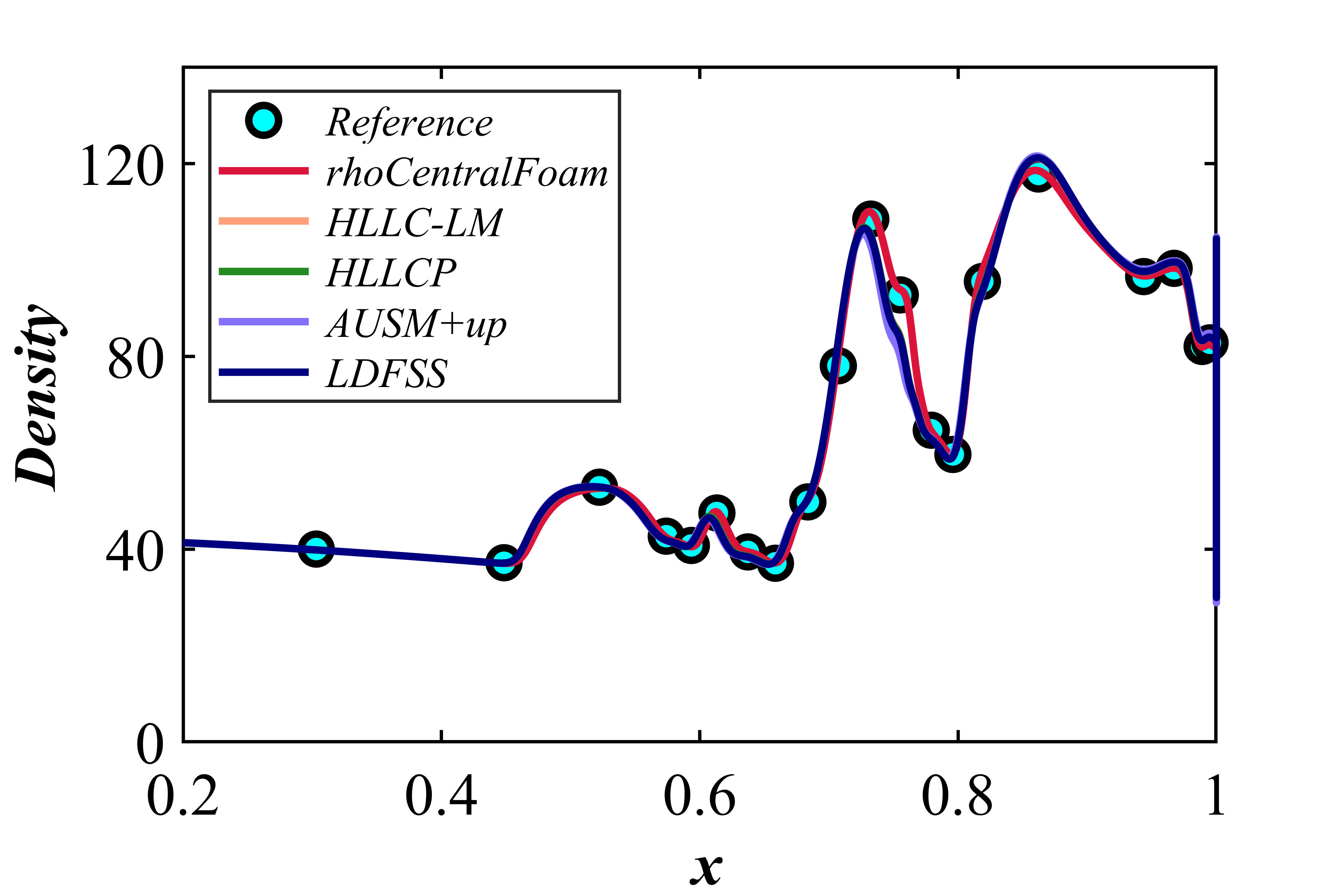}
\caption{Comparison of the density distribution over the bottom wall of the shock tube at $t = 1$ for the viscous shock tube problem at Reynolds number 200 using a fine grid obtained using different flux schemes. }\label{vs_re_200_fine_plot}
\end{figure}

Figure \ref{vs_re_200_coarse} shows the density distribution with the default solver in the OpenFOAM framework and the implemented schemes. The result of the default solver indicates that the resolution of the contact wave near the end of the tube is lost, while it is resolved better for the other schemes. Notably, the separation shock foot is steep and downstream. As a consequence, the separation bubble is shorter. The primary vortex is smeared along with the loss of resolution of the corner wall vortex. The HLLCP scheme in Figure \ref{vs_re_200_coarse} (c) generates a slightly oscillatory solution behind the reflected shock wave. The other flow features are marginally different in terms of engineering accuracy, and no other significant conclusions can be derived from these qualitative arguments.

Figure \ref{vs_re_200_coarse_plot} shows the density distribution over the bottom wall of the tube obtained with different schemes and is compared against the reference data of Daru et al. \cite{daru2009numerical} at selected points. The density rise at the separation location is delayed for the default solver as the lambda shock foot is placed downstream, corroborated in Figures \ref{Figure5} (a) and (b). The HLLCP scheme generates the highest peak near $x = 0.8$, suggesting it to be least dissipative at least in the separation zone. The HLLC, HLLCP, AUSM+up, and LDFSS schemes improve the resolution of the first peak; however, they are substantially underpredicted both for their magnitude and location. This is even worse for the default solver owing to its higher numerical dissipation.

\begin{figure}[h!]
\centering
\includegraphics [width=0.5\textwidth]{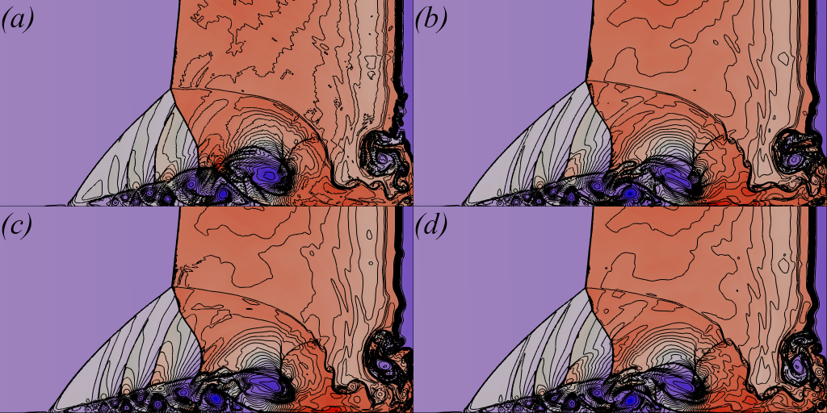}
\caption{Spatial density distribution at $t = 1$ for viscous shock tube problem at Reynolds number 2500 using a fine grid obtained using different flux schemes. (a) \textit{rhoCentralFoam} (b) HLLC-LM (c) HLLCP (d) AUSM+up}\label{vs_re_2500_fine}
\end{figure}

Figures \ref{vs_re_200_fine} and \ref{vs_re_200_fine_plot} show the density distribution contour and distribution over the bottom wall of the tube, respectively, using the fine mesh. Unsurprisingly, all the features' resolutions improve significantly, irrespective of the scheme employed. The primary vortex is significantly elongated in the transverse direction, suggesting a reduction in numerical dissipation owing to a reduction in the numerical truncation error. Notably, the location of the lambda shock foot in the default solver's solution agrees well with the other schemes used in this investigation, which re-confirms the dissipative nature of these schemes (i.e., other schemes performed better even on the significantly coarse mesh). The spurious solution of HLLCP diminished (by grid refinement) and is confined only to the small region behind the reflected shock. In contrast, the KNP scheme \textit{rhoCentralFoam} generate these numerical instabilities at the exact location on the fine mesh, which suggests that the scheme’s numerical behavior (dissipative and dispersive) is strongly dependent on the grid size, and the authors speculate that they may be amplified even more on refining the grid further and lead to solution convergence issues. Therefore, these numerical instabilities can contaminate the solution after a long duration of the simulation. Similarly, AUSM+up in Figure \ref{vs_re_200_fine} (e) introduces numerical instability, especially behind the reflected shock wave, for similar reasons to the one-dimensional shock tube problem. This pressure difference also manifests in the standing contact wave by producing instabilities over it with a shorter wavelength than other shock-capturing schemes. In this sense, HLLC and LDFSS generate the most ideal results on a fine grid in the presence of actual physical viscosity.

\begin{figure}[h!]
\centering
\includegraphics [width=0.5\textwidth]{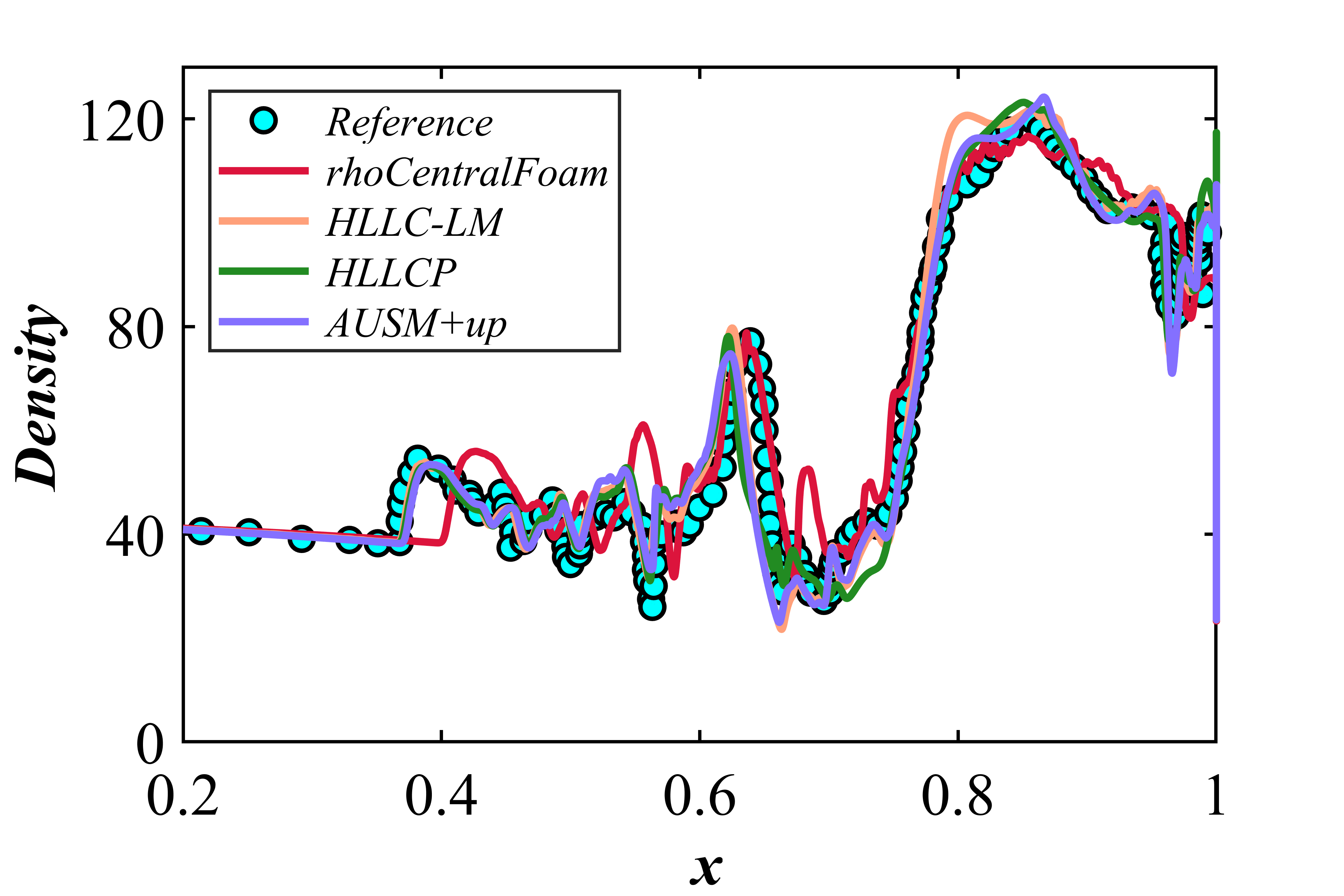}
\caption{Comparison of the density distribution over the bottom wall of the shock tube at $t = 1$ for the viscous shock tube problem at Reynolds number 2500 using a fine grid obtained using different flux schemes.}\label{vs_re_2500_fine_plot}
\end{figure}

Figure \ref{vs_re_200_fine_plot} represents the density distribution for the fine mesh. The first peak near $x = 0.6$ is well matched with the reference numerical results in magnitude and location, which is absent for the coarse grid simulation, irrespective of the schemes. The peaks are slightly sensitive to the scheme; however, the differences are insignificant to the extent of engineering accuracy. Overall, there is a decent agreement for all the schemes with the reference study on this extremely fine grid.

We extend this investigation to an extremely high Reynolds number on the fine grid, where the effects of real physical viscosity are diminished. This test is ideal for evaluating solvers/schemes' behaviour in terms of their stability and convergence characteristics. To facilitate this test, we set the Reynolds number at 2500 through the dynamic viscosity, as explained earlier in this section. Figure \ref{vs_re_2500_fine} shows these results. The default solver produces meaningful results, but the spurious oscillation in the post-shock region shows up along with a slightly smaller separated region compared to other schemes; nevertheless, all the other schemes provide consistent results, but the LDFSS scheme fails with floating-point error. The lower physical viscosity (high Reynolds number), coupled with a lower numerical viscosity (fine grid simulation), leads to its failure. The density distribution in Figure \ref{vs_re_2500_fine_plot} over the lower wall is consistent for all the numerics considered here, except that the solution of the default solver proceeds with \textit{wiggles}, and severe underprediction particularly in the post-shock region. The separation shock foot is located downstream compared to the reference solution; thus, the solution with the default solver shows incorrect locations of the primary flow features.

\subsection{Shock diffraction over wedge at $M_s = 1.3$}
The third problem of this series involves shock diffraction over a stationary wedge. The planar moving shock wave of strength $ M_s = 1.3$ impinges on the apex of the wedge, later reflecting as the bow shock wave as time progresses. This process also forms a Mach stem, which glides over the wedge's surface and ultimately diffracts around the corner, producing a supersonic vortex. This primary vortex is fed by the shear layer attached to the corner, and at the same time, the diffracted shock wave splits into accelerated and decelerated shock waves. The shear layer undergoes KHI, and tiny vortexlets orbit around the primary vortex core. Meanwhile, they also interact with the decelerated shock wave gliding in the opposite direction of the primary vortex core's rotation, thus producing Diverging Acoustic Waves (DAW). Therefore, we suitably use this case to understand how the schemes resolve these complicated flow field features, especially the DAW. The resolution of DAW finds its application in various aero-acoustic problems. 

\begin{figure}[h!]
\centering
\includegraphics [width=0.5\textwidth]{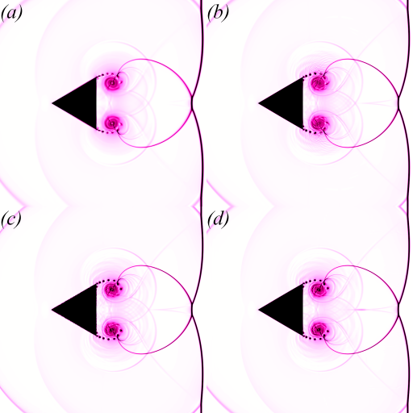}
\caption{Density gradient magnitude at $t_f = 128 \mu s $ for the shock diffraction over a wedge at $M_s = 1.3$ with different flux schemes. (a) \textit{rhoCentralFoam} (b) HLLC-LM (c) HLLCP (d) AUSM+up }\label{wedge}
\end{figure}

To simulate this numerically, we consider the numerical setup suggested by Kofoglu et al. \cite{kofoglu2022vortexlet}. The interested readers can refer to the computational domain used in their study. All the boundaries are treated as non-reflective, except that the bottom boundary is symmetric, as we simulate only that half portion of the wedge. We apply no-slip adiabatic wall conditions to the wedge. Real air properties are used to represent the specific heat and Prandtl number. The viscosity of the fluid is calculated using Sutherland's formulation. A non-stationary normal shock wave is initialised at the apex of the wedge, and we measure the time starting from this moment. We restrict the CFL number to 0.2 for this test. The reference solution is available in the experimental work of Chang and Chang \cite{chang2000shock}.

Figure \ref{wedge} shows these results in the form of numerical schlieren (or density gradient magnitude) with the reflected bottom half of the domain, obtained with different numerical methods at $t_f = 128 \mu s$. Once again, the default solver in Figure \ref{wedge} (a) resolves the primary flow features such as various shock waves, primary vortex core and vortexlets; however, the DAW is significantly dissipated, with just the head of the DAW visible. The corresponding shear layers through various triple points are also dissipated, suggesting amplified numerical viscosity of the default solver. On the other hand, the AUSM scheme in Figure \ref{wedge} (b) outperforms the default solver in terms of resolving these weak waves; however, we observe the flowfield to be slightly oscillatory. The HLLC-based scheme surpasses in terms of accurately resolving the tiny vortexlets and DAW without introducing any spurious oscillations. The LDFSS scheme is unstable and diverges during a few initial iterations while resolving the Mach stem before the diffraction phenomena.

\subsection{Ramp-induced shock wave boundary layer interaction at Mach 6}

\begin{figure}[h!]
\centering
\includegraphics [width=0.5\textwidth]{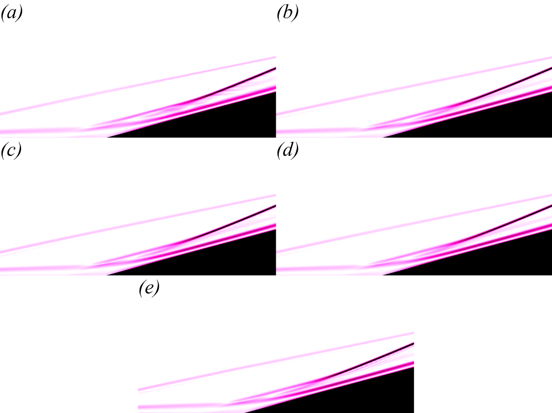}
\caption{Density gradient magnitude at steady state for the RSBLI problem at Mach 6 with different flux schemes. (a) \textit{rhoCentralFoam} (b) HLLC-LM (c) HLLCP (d) AUSM+up (e) LDFSS }\label{rsbli}
\end{figure}

We now consider the steady state problem of a ramp-induced shock wave boundary layer interaction at Mach 6. We use this case to assess the stability of the solver/scheme at high Mach numbers but considering the effects of physical viscosity. We use the same geometrical configuration with a wedge angle of $15^\circ$ as suggested by Marini \cite{marini1998effects} and the computational domain of John et al. \cite{john2014shock} A free stream of Mach 6 blows over this ramp-flat plate junction, generating complex shock interaction phenomena along with a separation bubble and its associated shock waves situated at the corner. We impose the Dirichlet boundary conditions at the inlet with the values $p_{\infty} = 199.435$ Pa, $T_{\infty} = 131.7$ K, and $U_{\infty} = 1380 $ m/s corresponding to the free-stream Mach number of 6. The wall is assumed to follow no-slip conditions maintained at wall temperature $T_w =300$ K. The rest of the boundaries are treated as non-reflecting. The domain is discretised with 420 cells in x-direction and 240 cells in y-direction, with the wall resolved at $5 \mu m$, ensuring $y^+ =1$. Properties of air are used to represent the specific heat, $C_p$ and Prandtl number, Pr. Sutherland's equation is used to represent the viscosity of the fluid. The flow is assumed to be laminar. We advance the solution with CFL = 0.5 to test the effects at higher Courant numbers but considering the physical viscosity, and seek the results at steady state obtained at $t_f = 1$ ms. 

\begin{figure}[h!]
\centering
\includegraphics [width=0.5\textwidth]{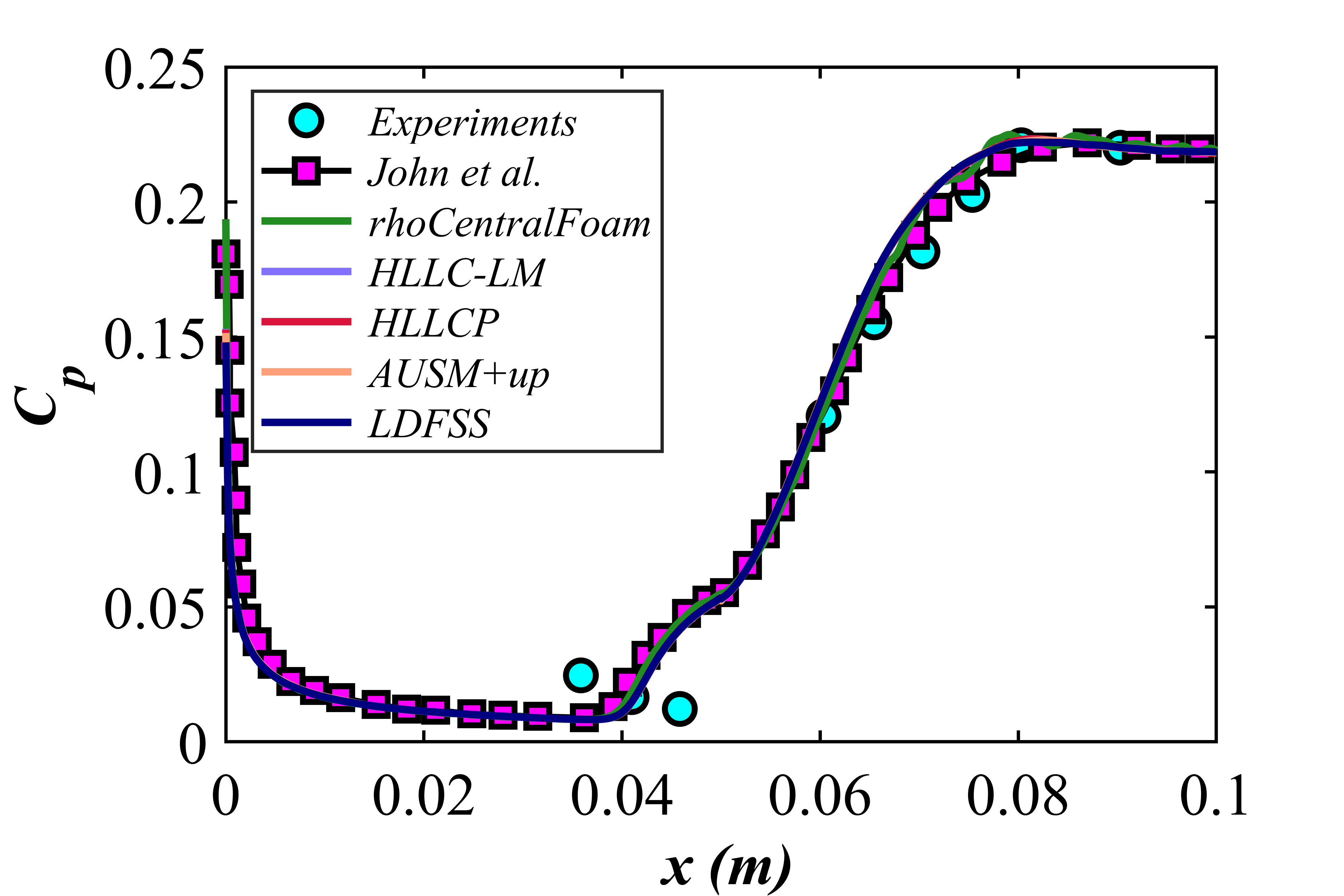}
\caption{Comparison of the pressure coefficient, $C_p$, over the flat plate-ramp surface at steady state for the RSBLI problem at Mach 6 obtained using different flux schemes.}\label{rsbli_plot}
\end{figure}

Figure \ref{rsbli} is the outcome of this test, represented as the numerical schlieren image. Figure \ref{rsbli} (a) corresponds to the results of the default solver, which shows a high level of numerical oscillation, especially behind the separated shock wave. All the other schemes generate comparable results, even at the CFL number of 0.5, showing their excellent robustness even in the hypersonic regime. However, the LDFSS in Figure \ref{rsbli} (e) scheme shows these unwanted oscillations like the default solver behind the separated shock wave, which suggests that it is dispersive at high Mach numbers and may pose stability issues for the unsteady problems in the hypersonic regime. Figure \ref{rsbli_plot} compares the pressure coefficient $C_p$ obtained in the present investigation against the experimental and computational results of Marini \cite{marini1998effects} and John et al. \cite{john2014shock}, respectively. All the solvers match the reference results; however, we observe small-scale undulation in the distribution post the boundary layer's reattachment location corresponding to the default solver.


\subsection{Turbulent flow in hypersonic intake}

\begin{figure}[h!]
\centering
\includegraphics [width=0.5\textwidth]{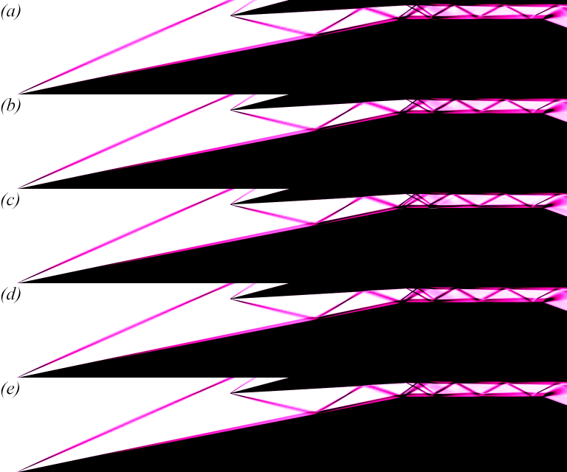}
\caption{Density gradient magnitude at steady state for the turbulent flow in hypersonic intake at Mach 4 at steady state with different flux schemes. (a) \textit{rhoCentralFoam} (b) HLLC-LM (c) HLLCP (d) AUSM+up (e) LDFSS }\label{emami}
\end{figure}

\begin{figure}[h!]
\centering
\includegraphics [width=0.5\textwidth]{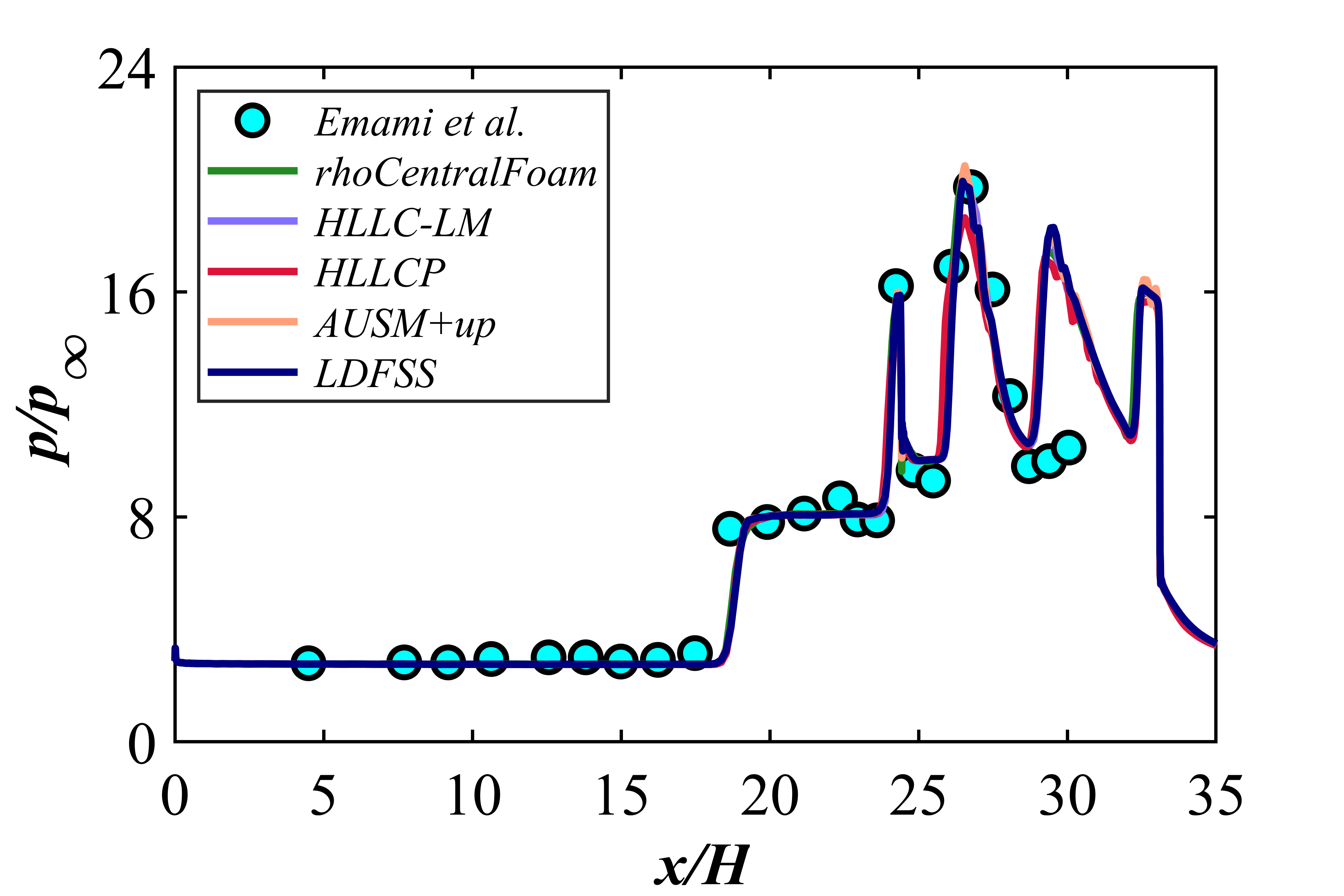}
\caption{Comparison of the normalized pressure, $p/p_{\infty}$, over the floor wall of isolator for the turbulent flow in hypersonic intake at Mach 4 at steady state obtained using different flux schemes. }\label{emami_plot}
\end{figure}

The last problem in this series is the simulation corresponding to the turbulent flow inside the hypersonic intake. This test case allows us to show the characteristics of the scheme in the presence of turbulence modelling, where the turbulence model's viscosity can play a crucial role in the stability and accuracy of the schemes. To facilitate this comparison, we simulate the two-dimensional flow inside the scramjet intake. The intake model resembles the several intake configurations tested by Emami et al. \cite{emami1995experimental} using experimental aid. Interested readers can refer to the report for the detailed design of the intake model. We use the long cowl, long isolator configuration with cowl opening angle, $\beta = 6^\circ$. The author used the tip-to-tail configuration; however, we restrict the simulations to the intake fore-body, the isolator section, and a slightly extended portion of the expansion section. The corresponding computational domain is given in the Appendix \ref{app1} of this article. The computational domain is discretised with 205,000 cells, with the first cell height of $50 \mu m$. The inlet is imposed with Dirichlet conditions corresponding to a Mach number of 4. The concerning free stream total pressure and temperature are $p_0 = 1.31$ MPa and $T_0 =  288$ K respectively. All the other boundaries are imposed with supersonic outflow conditions, except that the wall is assumed to be no-slip with insulated conditions for temperature. We rely on the standard wall functions to resolve the turbulent boundary layer. We restrict the CFL number to 0.2 and seek the mean solution at steady state. SST $k-\omega$ model is used to model turbulence where the turbulent kinetic energy, k, is calculated by assuming the free stream turbulence intensity of $5\%$, and the dissipation scale is calculated by using the throat height of the intake as the length scale.

Figure \ref{emami} shows the numerical schlieren images for all the numerics used in this investigation. All the schemes, including the default solver, show physically consistent results with negligible variation in the reflected shock system inside the isolator concerning the shock wave locations. All the solutions are devoid of any numerical noise, suggesting that the mean flow resolved using the turbulence model is insensitive to the flux scheme used in this study. Figure \ref{emami_plot} shows the pressure coefficient on the bottom wall of the intake fore-body and isolator, which shows these comparisons quantitatively. The pressure distribution is slightly underpredicted at the highest peak pressure location for the HLLCP scheme. In contrast, the other scheme's results are decently comparable with the experimental readings of Emami et al. \cite{emami1995experimental} The AUSM+up and LDFSS scheme shows minute oscillations in the pressure distribution.

\section{Conclusion and Summary\label{Section5}}

\begin{figure*}[t]
\centering
\includegraphics [width=1\textwidth]{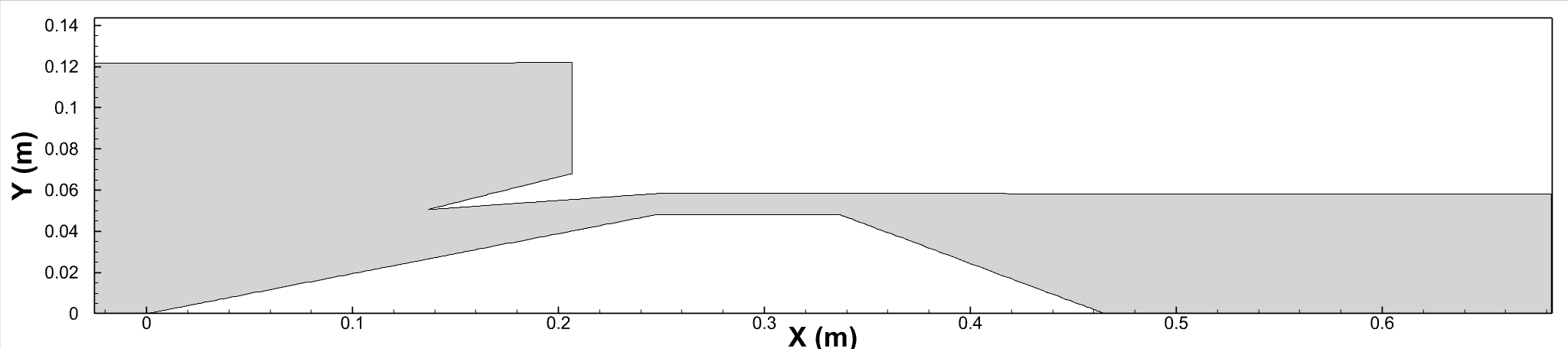}
\caption{Computational domain for the turbulent flow in hypersonic intake at Mach 4. }\label{domain}
\end{figure*}

The novelty of this article resides in showing how to numerically implement contemporary shock-capturing schemes and enhance the predictor-corrector algorithm used in the default density-based solver in the OpenFOAM framework. These methods are then tested for a variety of canonical cases of increasing complexity, ranging from theoretical problems to problems of industry caliber. The article's first section primarily focuses on how the implemented schemes/algorithms enhance the default solver's dissipation and dispersion properties. The default solver appears to be highly diffusive on a coarse grid with few cells, according to preliminary tests using the Euler equation in one dimension. However, when the fidelity is increased by increasing the grid points, the default solver's solutions become unstable and contaminate the solution. This is mainly the result of solving the governing equations in the successive steps (see algorithm), where the updated velocity field in the current time step is used to calculate the temperature (through total energy) using the total energy from the previous time iterations. In essence, this lag in updating the temperature field accumulates numerical errors over time, which grow substantially towards the end of the simulation. However, these numerical noises are significantly reduced by the enhanced predictor-corrector technique that makes use of the TVD-based Runge-Kutta scheme. By self-adjusting the residuals in the internal iterations of the Runge-Kutta steps, the TVD-based algorithm's monotone-preserving property limits the growth of spurious oscillation.

Using a high CFL number up to 0.5 on a fine grid without creating these spurious oscillations is another benefit of employing this better time integration. Although this CFL number is supported by the default solver based on the central schemes of KT/KNP schemes, the usage of the first-order Euler time integration for the predictor-corrector technique alone may result in significant convergence problems on the fine grid. We hypothesize that the numerical stability attained could be ruined by adjusting the Runge-Kutta coefficients corresponding to Euler integration. On the other hand, by using the enhanced predictor-corrector approach, the default solver with KNP and KT scheme can be improved; however, we do not specifically demonstrate these results in this work.


The second part of this article further points to the dissipation and dispersion characteristics and the underlying stability and accuracy of the scheme by solving the Euler and Navier-Stokes equations for one-dimensional to three-dimensional problems in varied Mach number regimes. As far as the Riemann solvers are concerned, the improved versions of the HLLC scheme, i.e., HLLC-LM and HLLCP schemes, show excellent robustness across the high Mach number flows (transonic and supersonic regimes), both using the structured and unstructured grid. It is well known that the schemes, which explicitly resolve the contact discontinuity, often suffer from the grid-aligned instability, i.e, when a strong discontinuity aligns perfectly with the grid lines of the mesh employed (e.g. see the case of strong shock diffracting around the corner or Sedov blast wave or steady shock wave in front of a cylinder) \cite{dumbser2004matrix}. As stated earlier in this article, they are the artefacts of vanishing numerical flux in the transverse direction and subsequently the incorrect scaling of the acoustic speed. The HLLC-LM and HLLCP schemes cure these unwanted numerical errors that are generated in the post-shock region, while simultaneously preserving the contact discontinuities. It is further noted that the HLLCP is a clear winner in the umbrella of the HLLC-type schemes, with HLLC-LM a slightly inferior scheme under the low Mach number limit, e.g., see Taylor Green Vortex flow or the incompressible flow over a cylinder. 

The flux vector splitting schemes (AUSM+up and LDFSS), which separately treat the convective and pressure fluxes, show slightly poor resolution of the contact discontinuities, e.g., see the modified shock tube problem, the muzzle blast problem, or shock diffraction over the wedge. Although the LDFSS scheme outperforms the AUSM+up scheme under a low Mach number limit (e.g., see Double shear-layer problem, Taylor-Green vortex decay problem or incompressible flow over a cylinder), it outrightly fails to provide stable solutions for problems involving strong moving discontinuities, as demonstrated for the strong shock diffracting around a corner or a wedge, Sedov blast wave and Mach 20 flow over the cylinder, indicating its poor dissipation characteristics for the strong shock system. This scheme even fails to provide results under the limit of high Reynolds number flow on a refined grid, such as Re 2500 flow in the viscous shock tube. It is also noted that the LDFSS scheme is expensive in terms of iterations involved to resolve the flow in the low Mach number limit, as the stable solutions are obtained on CFL numbers which are at least two orders of magnitude smaller than the CFL numbers used for other schemes. The comparison of these schemes with the high fidelity solver \textit{rhoEnergyFoam} reveals HLLCP to possess comparable resolution characteristics to all the discontinuities present in the system, if not better, for the entire range of Mach numbers and Reynolds numbers reported in this piece of work.

	\begin{acknowledgments}
		The authors would like to acknowledge the National Supercomputing Mission (NSM) for providing computing resources of “Param Sanganak” at IIT Kanpur, which is implemented by CDAC and supported by the Ministry of Electronics and Information Technology (MeitY) and the Department of Science and Technology (DST), Government of India. We would like to acknowledge the IIT-K Computer Centre (www.iitk.ac.in/cc) for providing the resources to perform the computational work. This support is gratefully acknowledged.
	\end{acknowledgments}
	
	
	\section*{AUTHOR DECLARATIONS}
	
	\subsection*{Conflict of Interest}
	The authors have no conflicts to disclose.
	
	\section*{Data Availability}
	
	The data that support the findings of this study are available from the corresponding author upon reasonable request.

\section*{Funding sources}
Financial support for this research is provided through the DIA-CoE scheme of DRDO, India.

\appendix

\section{\label{app1} Computational domain for the turbulent flow inside a scramjet intake at Mach 4}

The Figure \ref{domain} shows the computational domain for the turbulent flow inside a scramjet intake at Mach 4.

	\section*{REFERENCES}
	
	\nocite{*}
	\bibliography{aipsamp}
	
\end{document}